\documentclass{jfm} \usepackage{graphicx} \usepackage{natbib}
\usepackage{amsbsy} \usepackage{amsmath} \usepackage{amssymb}
\usepackage{bm} \usepackage{color}

\usepackage[T1]{fontenc}
\usepackage[utf8]{inputenc}
\usepackage{bm}
\usepackage{authblk}
\usepackage[percent]{overpic} 
\usepackage{subfigure}
\usepackage{color}

\title[Global and local statistics in turbulent convection at low Prandtl numbers]{Global and local statistics in turbulent convection at low Prandtl numbers}
\author[1]{Janet D. Scheel}
\author[2]{J\"org Schumacher}
\affil[1]{\small{Department of Physics, Occidental College, 1600 Campus Road, M21, Los Angeles, California 90041, USA}} 
\affil[2]{\small{Institut f\"ur Thermo- und Fluiddynamik, Postfach 100565, Technische Universit\"at Ilmenau, D-98684 Ilmenau, Germany}}
                  
\date{\today}

\begin{document}
\maketitle

\abstract{Statistical properties of turbulent Rayleigh-B\'{e}nard convection at low Prandtl numbers $Pr$, which are typical for liquid 
metals such as mercury or gallium ($Pr\simeq 0.021$) or liquid sodium ($Pr\simeq 0.005$), are investigated in high-resolution 
three-dimensional spectral element simulations in a closed cylindrical cell with an aspect ratio of one and  are compared to 
previous turbulent convection simulations in air for $Pr=0.7$. We compare the scaling of global momentum and heat transfer. 
The scaling exponent $\beta$ of the power law $Nu=\alpha Ra^{\beta}$ is  $\beta=0.265\pm 0.01$ for $Pr=0.005$ and  
$\beta=0.26\pm 0.01$ for $Pr=0.021$, which are smaller than that for convection in air ($Pr=0.7$, $\beta=0.29\pm 0.01$). 
These exponents are in agreement with experiments. Mean profiles of the root-mean-square velocity as well as the thermal and kinetic energy dissipation rates 
have growing amplitudes with decreasing Prandtl number which underlies a more vigorous bulk turbulence in the low--$Pr$ 
regime. The skin-friction coefficient displays a Reynolds-number dependence that is close to that of an isothermal, intermittently 
turbulent velocity boundary layer. The thermal boundary layer thicknesses are larger as $Pr$ decreases and conversely the 
velocity boundary layer thicknesses become smaller. We investigate the scaling exponents and find a slight decrease in exponent magnitude 
for the thermal boundary layer thickness as $Pr$ decreases, but find the opposite case for the velocity boundary layer thickness 
scaling.  A growing area fraction of turbulent patches close to the heating and cooling plates can be detected by exceeding a 
locally defined shear Reynolds number threshold. This area fraction is larger for lower $Pr$ at the same $Ra$, but the scaling 
exponent of its growth with Rayleigh number is reduced. Our analysis of the kurtosis of the locally defined shear Reynolds 
number demonstrates that the intermittency in the boundary layer is significantly increased for the lower Prandtl number and for sufficiently high
Rayleigh number compared to convection in air. This complements our previous findings of enhanced bulk intermittency in low-Prandtl-number
convection.}

\section{Introduction}
Many turbulent thermal convection phenomena in nature and technology are present in low--Prandtl--number fluids for which 
the  kinematic viscosity of the fluid, $\nu$, is much smaller than the thermal diffusivity of the temperature field, $\kappa$, or 
in other words for which the ratio of both, the Prandtl number  
\begin{equation}
Pr=\frac{\nu}{\kappa}\,.
\end{equation}  
is much smaller than one. Applications range from turbulent convection in the Sun (\cite{Hanasoge2015}) or the liquid metal 
core of the Earth (\cite{King2013}) to nuclear engineering (\cite{Groetzbach2013}) or liquid metal batteries (\cite{Kelley2014}). 
In all these examples $Pr\lesssim 2\times 10^{-2}$, and in the solar convection case it could be smaller than $Pr\sim 10^{-4}$. 
Convective turbulence in these examples appears mostly in combination with other physical processes, such as rotation, radiation 
or the generation of magnetic fields.  Rayleigh-B\'{e}nard convection (RBC), a fluid flow in a layer which is cooled from above and 
heated from below, is the simplest example of a turbulent convective flow.

The record of experimental and numerical data for low-Prandtl-number convection is much smaller than for convection in air or water. 
Laboratory experiments are challenging in several ways. For Prandtl numbers $Pr<0.2$, opaque liquid metals such 
as liquid mercury or gallium at $Pr\simeq 0.025$ (\cite{Rossby1969,Cioni1996,Cioni1997,Takeshita1996,Glazier1999,Aurnou2001,King2013}) 
or liquid sodium at $Pr\simeq 0.005$ (\cite{Horanyi1999,Cioni1997a,Frick2015}) have to be used. Additionally, experiments in 
sodium require working temperature of more than 100$\,^{\circ}$C. An optical access to the flow by laser-imaging techniques is 
impossible. The heating and cooling at the bottom and top walls is usually established with copper plates. In liquid sodium experiments, 
the thermal conductivity of the plates is thus not significantly higher than that of the working fluid. This introduces significant 
variations of the temperature at the plates and can alter the magnitude of the heat transfer since the boundary conditions 
will be a mixture of prescribed temperature and prescribed normal temperature derivative, known as Robin boundary conditions.
For an experimental analysis we refer to \cite{Horanyi1999}. In low-Prandtl number convection the thermal boundary layer at the 
solid heating and cooling plates is much thicker than the velocity boundary layer. This has important implications for the turbulent heat and momentum transfer which is one focus of the research in turbulent convection (e.g. in \cite{Grossmann2000,Chilla2012, Scheel2014}). The global heat transfer is quantified by the dimensionless Nusselt number, $Nu$, while the global momentum transfer is measured by the Reynolds number, $Re$.  Detailed definitions of both parameters 
are given in section 2.1. Both the Nusselt and Reynolds numbers are functions of the Prandtl number $Pr$ and a second dimensionless 
parameter which quantifies the thermal driving of the turbulence, namely the Rayleigh number $Ra$. Different scaling laws $Nu(Ra)$
have been reported in different convection experiments for mercury and gallium, i.e., at fixed Prandtl number. In liquid sodium, only one experimental data record by \cite{Horanyi1999} exists which is similar to the parameters in this paper. This discussion implies that direct numerical simulations (DNS) are important for gaining access to the full three-dimensional turbulent fields in low-$Pr$ convection and to assure the accuracy of the boundary conditions. 

In the present work, we study the local and global statistical properties of turbulent convection flows at very low Prandtl numbers as present 
in liquid mercury and liquid sodium. We analyze a series of very high-resolution DNS obtained for moderate Rayleigh numbers in a cylindrical cell 
with an aspect ratio of one. We obtain global scaling laws for the turbulent heat and momentum transfer and compare our findings with available 
data as  well as the scaling theory of \cite{Grossmann2001} and a recent update by \cite{Stevens2013}. In addition, we analyze the turbulence 
statistics for a series of three DNS runs which have been conducted at different Rayleigh and Prandtl numbers but with the same Grashof number 
$Gr=Ra/Pr$ for all three runs (for definition see section 2.2).  This perspective (\cite{Schumacher2015}) leaves the dimensionless Navier-Stokes equations 
unchanged and thus provides a more consistent comparison of convective flow at different Prandtl numbers from the point of view of fluid turbulence. 

A second focus of the present work is a detailed local statistical analysis of the dynamics in the boundary layers of velocity and temperature fields.   
While the heat transport is reduced in low--$Pr$ convection, production of vorticity and shear are strongly enhanced, both in the bulk and in the 
boundary layers. This is the reason why simulations of turbulent convection at very low Prandtl numbers become very demanding, as is reported 
for homogeneous turbulent flows (\cite{Mishra2010}), in planar turbulent flows between parallel free-slip or no-slip walls 
(\cite{Kerr2000,Breuer2004,Verma2012,Petschel2013}) or in closed cylindrical cells (\cite{Camussi1998,VanderPoel2013,Schumacher2015}). We 
 focus our interest on the boundary layer dynamics and determine that the enhanced intermittency in the bulk is connected to a stronger intermittent 
behavior in the boundary layer, in particular for the velocity field.  But, while the vorticity and shear are enhanced in  low Prandtl number convection, 
we also find that the scaling of many diagnostic quantities is weaker. We postulate that this is due to the reduced heat transport and weaker dependence 
of the Nusselt number on the Rayleigh number.

The outline of the paper is as follows. In the next section, we present the model equations, the numerical method and the data sets. The third 
section discusses the scaling laws for global heat and momentum transfer.  Section 4 turns to the statistics at constant Grashof number and is 
followed by a detailed boundary-layer analysis in section 5. Finally we summarize our findings and give a brief outlook for future work.     

\section{Numerical model}
\subsection{Equations of motion}
We solve the three-dimensional equations of motion in the Boussinesq approximation. The equations are made dimensionless by using
height of the cell $H$, the free-fall velocity $U_f=\sqrt{g \alpha \Delta T H}$ and the imposed temperature difference $\Delta T$. The equations
contain the three control parameters: the Rayleigh number $Ra$, the Prandtl number $Pr$ and the aspect ratio $\Gamma=D/H$ with the 
cell diameter $D$. The equations are given by
\begin{eqnarray}
\label{ceq}
 {\bm \nabla}\cdot {\bf u}&=&0\,,\\
\label{nseq}
\frac{\partial{\bf u}}{\partial  t}+({\bf u}\cdot{\bm\nabla}){\bf u}
&=&-{\bm \nabla}  p+\sqrt{\frac{Pr}{Ra}} {\bm \nabla}^2{\bf u}+  T {\bf e}_z\,,\\
\frac{\partial  T}{\partial  t}+( {\bf u}\cdot {\bm \nabla})  T
&=&\frac{1}{\sqrt{Ra Pr}} {\bm \nabla}^2  T\,,
\label{pseq}
\end{eqnarray}
where
\begin{equation}
Ra=\frac{g\alpha\Delta T H^3}{\nu\kappa}\,,\;\;\;\;\;\;\;\;Pr=\frac{\nu}{\kappa}\,.
\end{equation}
The free-fall Reynolds number $Re_f$ is given by 
\begin{equation}
Re_f=\frac{U_f H}{\nu}=\sqrt{\frac{Ra}{Pr}}\,. 
\end{equation}
The variable $g$ stands for the  acceleration due to gravity, $\alpha$ is the thermal expansion coefficient, $\nu$ is the kinematic viscosity, and 
$\kappa$ is thermal diffusivity. We use an aspect ratio of  $\Gamma=1$ here. No-slip boundary conditions for the fluid (${\bf u}=0$)  are applied at 
the walls. The side walls are thermally insulated ($\partial T/\partial {\bf n}=0$) and the top and bottom plates are held at constant dimensionless
 temperatures $T=0$ and 1, respectively. In response to the input parameters $Ra$, $Pr$ and $\Gamma$,  turbulent heat and momentum fluxes 
 are established. The turbulent heat transport is determined by the Nusselt number which is defined as  
\begin{equation}
Nu=1+\sqrt{Ra Pr}\langle  u_z  T\rangle_{V,t}\,.
\label{Nusselt}
\end{equation}
The turbulent momentum transport is expressed by the Reynolds number which is defined as
\begin{equation}
Re=u_{rms,V}\sqrt{\frac{Ra}{Pr}}=u_{rms,V}Re_f\;\;\;\;\text{with}\;\;\;\;u_{rms,V} =\sqrt{\langle u_i^2\rangle_{V,t}}\,,
\label{Reynolds}
\end{equation}
In both definitions $\langle\cdot\rangle_{V,t}$ stands for a combined volume and time average.

The equations are numerically solved by the Nek5000 spectral element method package which has been 
adapted to our problem. The code employs second-order time-stepping, using a backward difference formula. The whole set of
equations is transformed into a weak formulation and discretized with a particular choice of spectral basis functions (\citeauthor{Fischer1997} 
\citeyear{Fischer1997}, \citeauthor{Deville2002} \citeyear{Deville2002}). The resulting linear, symmetric Stokes problem is solved implicitly. 
This system is split, decoupling the viscous and pressure steps into independent symmetric positive definite subproblems which are solved 
either by Jacobi (viscous) or multilevel Schwartz (pressure) preconditioned conjugate gradient iteration. Fast parallel solvers based on 
direct projection or more scalable algebraic multigrid are used for the coarse-grid solve that is part of the pressure preconditioner.
For further numerical details and comprehensive tests of the sufficient spectral resolution, we refer to detailed investigations in \cite{Scheel2013}. 

\begin{figure}
\centering
\includegraphics[width=0.8\textwidth]{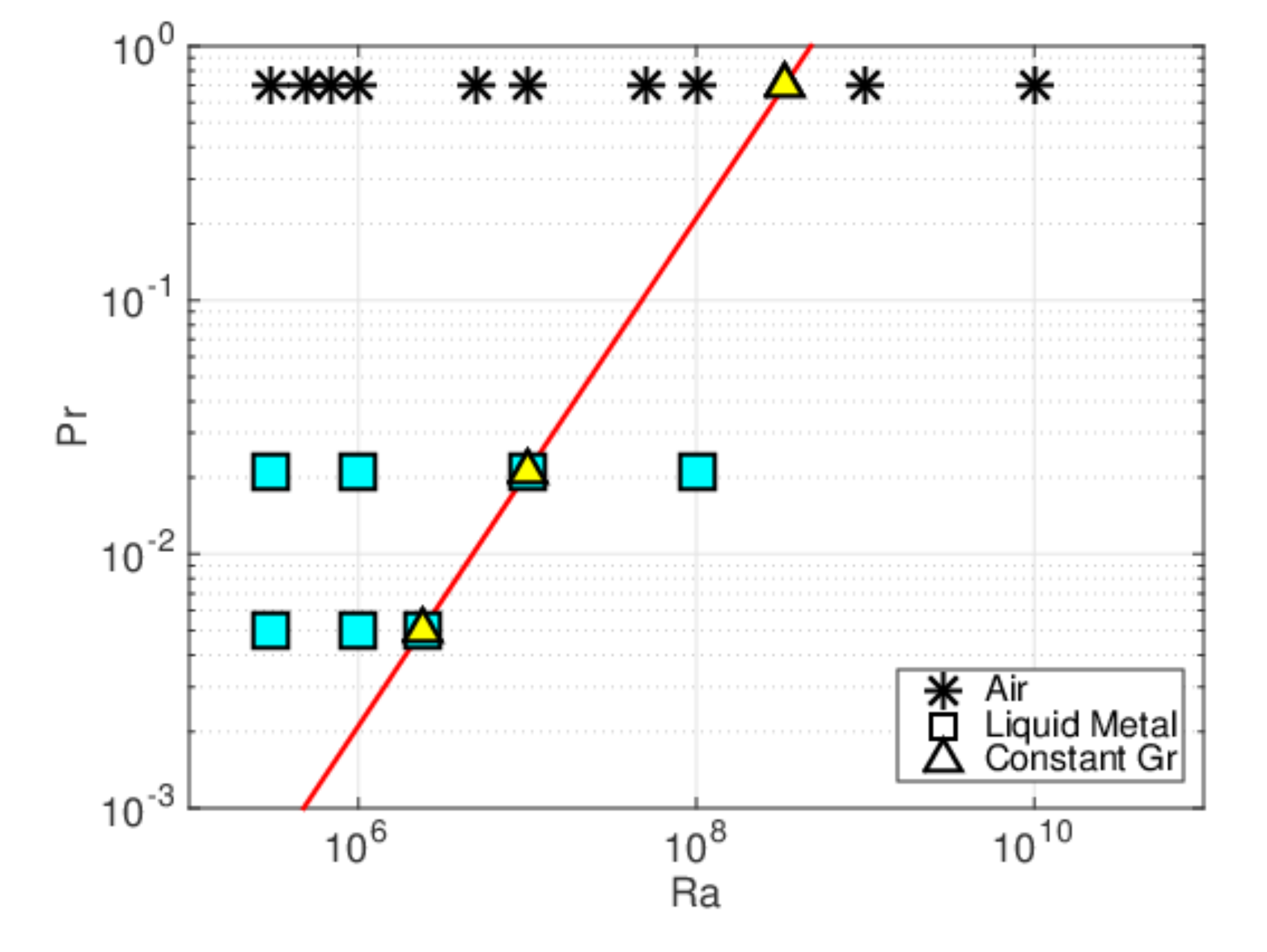}
\caption{(Colour online) Prandtl number-Rayleigh number parameter space. The solid diagonal line corresponds to data at constant Grashof number.
The different working fluids are indicated in the legend. The data for air are from \cite{Scheel2014}.}
\label{ravspr}
\end{figure}

\begin{table}
\begin{center}
\begin{tabular}{cccccccccc}
Run & $Ra$ &  $Pr$ & $N_{e}$ & $N$  & $N_{bl}$ & $Nu$ & $\delta_T/H$ & $u_{rms}$ & $Re$\\ \\
1 & $3.33\times 10^8$ & 0.7 & 875,520 & 11 & 40 & $44.9 \pm 1.6$ &  0.011 & $0.171\pm 0.003$  & $3720\pm 60$ \\
2 & $3\times 10^5$  & 0.021 & 256,000 & 7 & 54 & $4.29\pm 0.12$  & 0.117 & $0.483\pm 0.009$  & $1830\pm 30$\\
3 & $10^6$  & 0.021 & 256,000 & 9 & 48 & $5.43\pm 0.03$ & 0.093 & $0.439\pm 0.006$ & $3030 \pm 40$\\
4 & $10^7$  & 0.021 & 875,520 & 11 & 26 & $10.11\pm 0.05$ & 0.050 & $0.387\pm 0.005$ & $8450 \pm 100$\\
5 & $10^8$  & 0.021 & 2,374,400 & 13 & 18 & $19.1\pm 1.3$ & 0.026 & $0.332\pm 0.004$ & $22900 \pm 300$\\
6 & $3\times 10^5$  & 0.005 & 256,000 & 9 & 43 & $3.26\pm 0.02$ & 0.153  & $0.597\pm 0.003$ & $4620\pm 20$\\
7 & $10^6$  & 0.005 & 875,520 & 9 & 39  & $4.45 \pm 0.07$ & 0.112 & $0.586\pm 0.003$ & $8290\pm 40$\\
8 & $2.38\times 10^6$  & 0.005 & 2,374,400 & 11 & 30 & $5.66\pm 0.09$ & 0.088  & $0.590\pm 0.006$  & $12800\pm 120$\\
\end{tabular}  
\caption{Parameters of the different spectral element simulations. We show the Rayleigh number $Ra$, the Prandtl number $Pr$, 
the total number of spectral elements $N_e$, the polynomial order $N$, and the number of grid points inside the velocity boundary 
layer, $N_{bl}$. The latter is calculated by (\ref{velBL1}). We also show the Nusselt number $Nu$ (see (\ref{Nusselt})), 
the thermal boundary layer thickness $\delta_T$ (see (\ref{TBL})), the root-mean-square velocity ${u}_{rms}$, and the Reynolds 
number $Re$ (see (\ref{Reynolds})). The error bars  have been obtained by evaluating the results over the first and second 
halves of the corresponding  data set separately and taking the difference of both results.}
\label{Tabpran}
\end{center}
\end{table}

\subsection{Data sets}
The $Pr-Ra$ parameter space covered by the data in table \ref{Tabpran} is shown schematically  in figure \ref{ravspr}. 
We have a fairly complete data set for thermal convection in air at $Pr=0.7$ (see \cite{Scheel2014}), and have added  new data at the very low 
Prandtl numbers for liquid mercury, $Pr=0.021$, and liquid sodium, $Pr=0.005$. We have chosen the Rayleigh numbers in data sets 1, 4 and 8
of table \ref{Tabpran}  to keep the Grashof number $Gr$, defined as
\begin{equation}
Gr=\frac{g\alpha\Delta T H^3}{\nu^2}=\frac{Ra}{Pr}=Re_f^2\,,
\end{equation}
constant at $4.76\times 10^8$. Consequently, $Gr=Re^2/u_{rms,V}^2$ with equation (\ref{Reynolds}).

\begin{figure}
\centering
\includegraphics[width=0.5\textwidth]{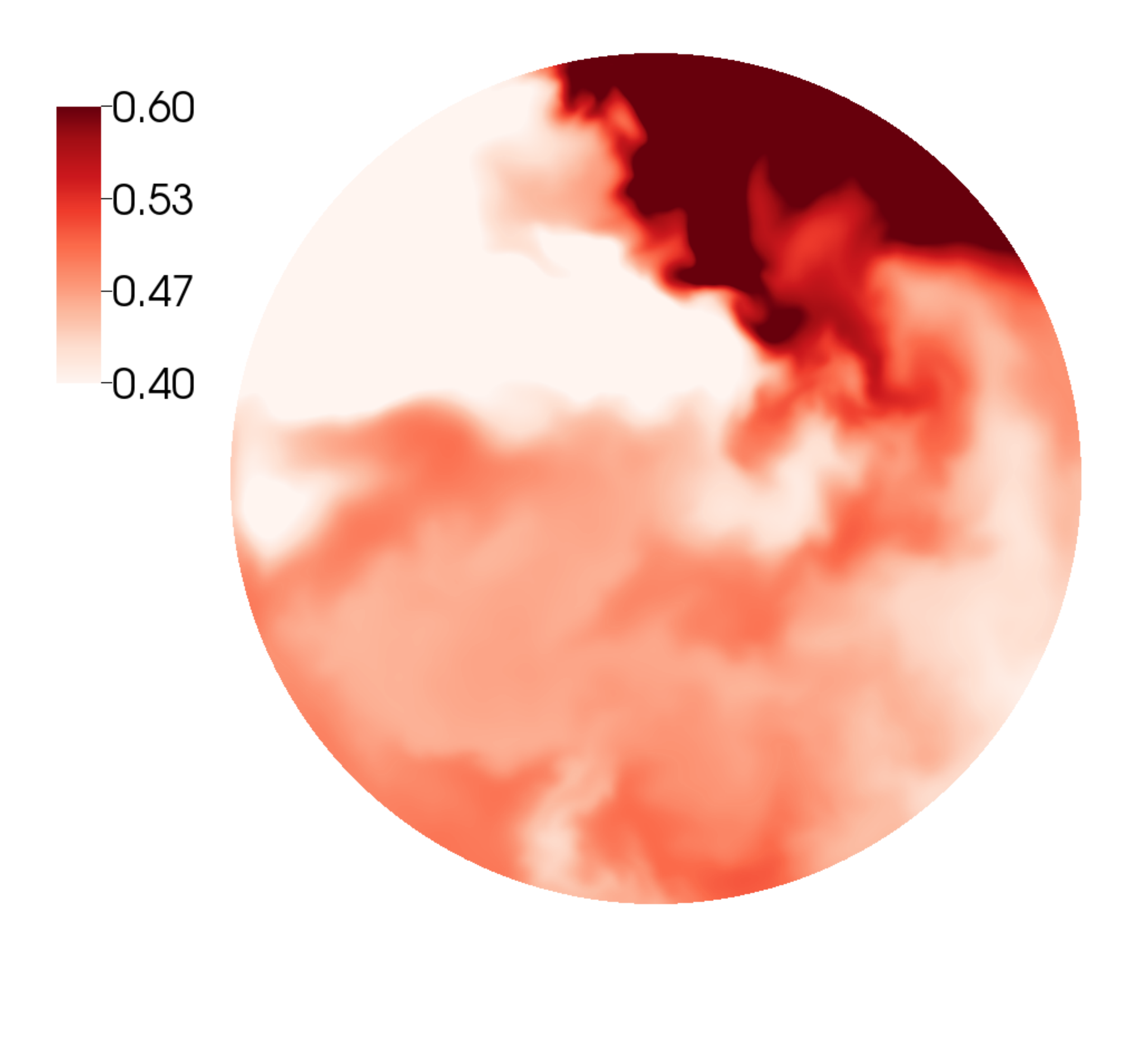}\includegraphics[width=0.5\textwidth]{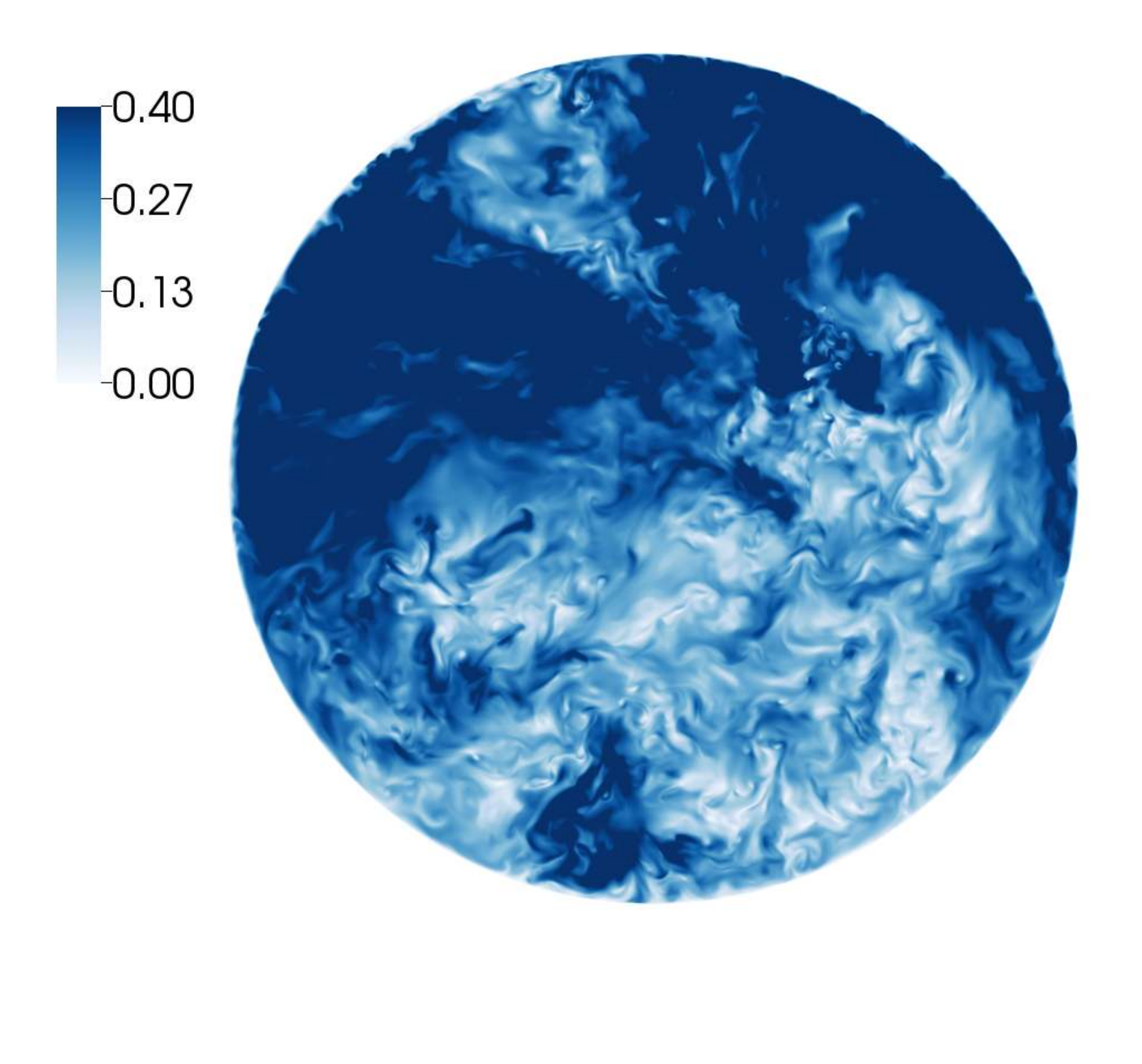}
\includegraphics[width=0.5\textwidth]{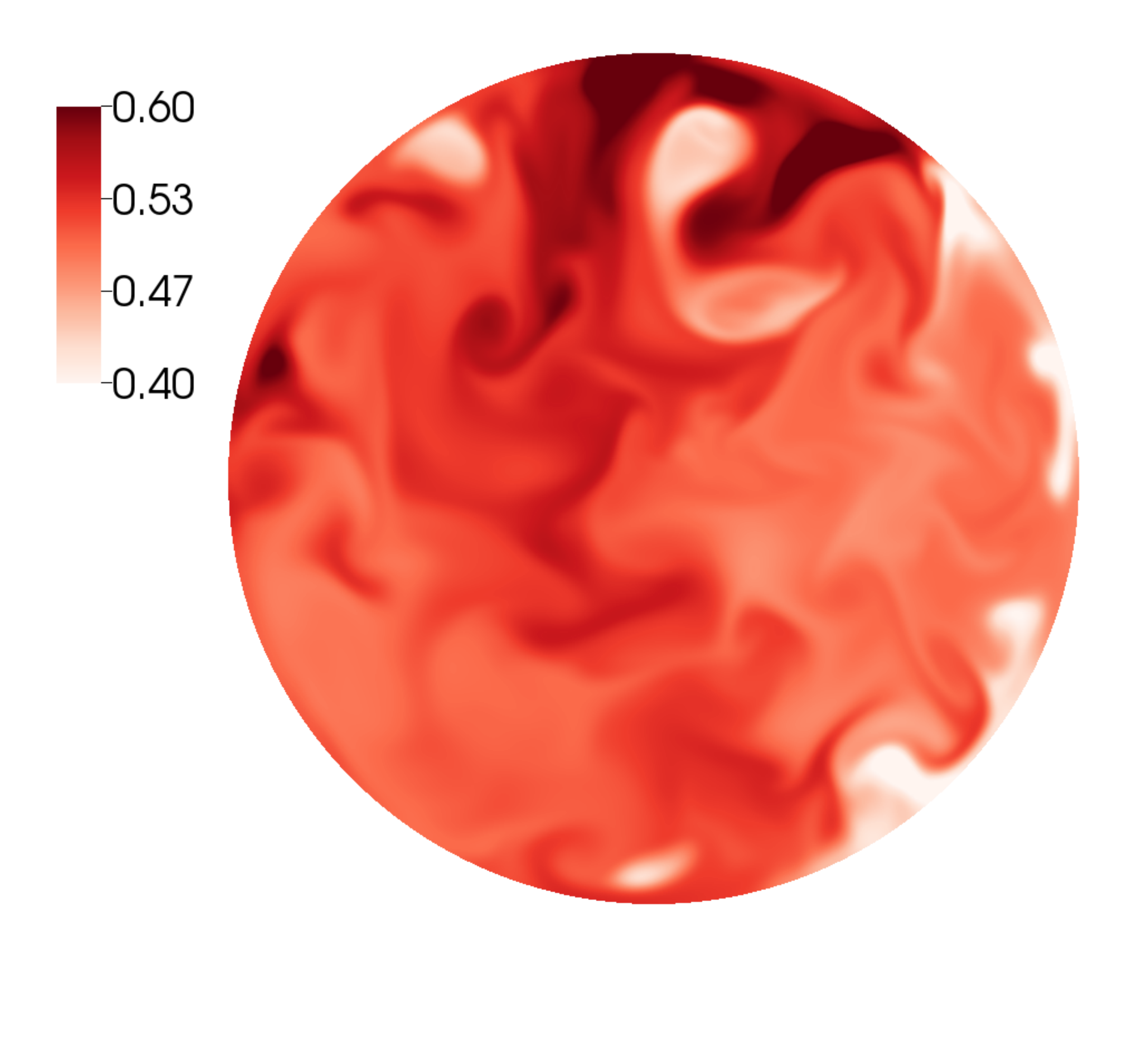}\includegraphics[width=0.5\textwidth]{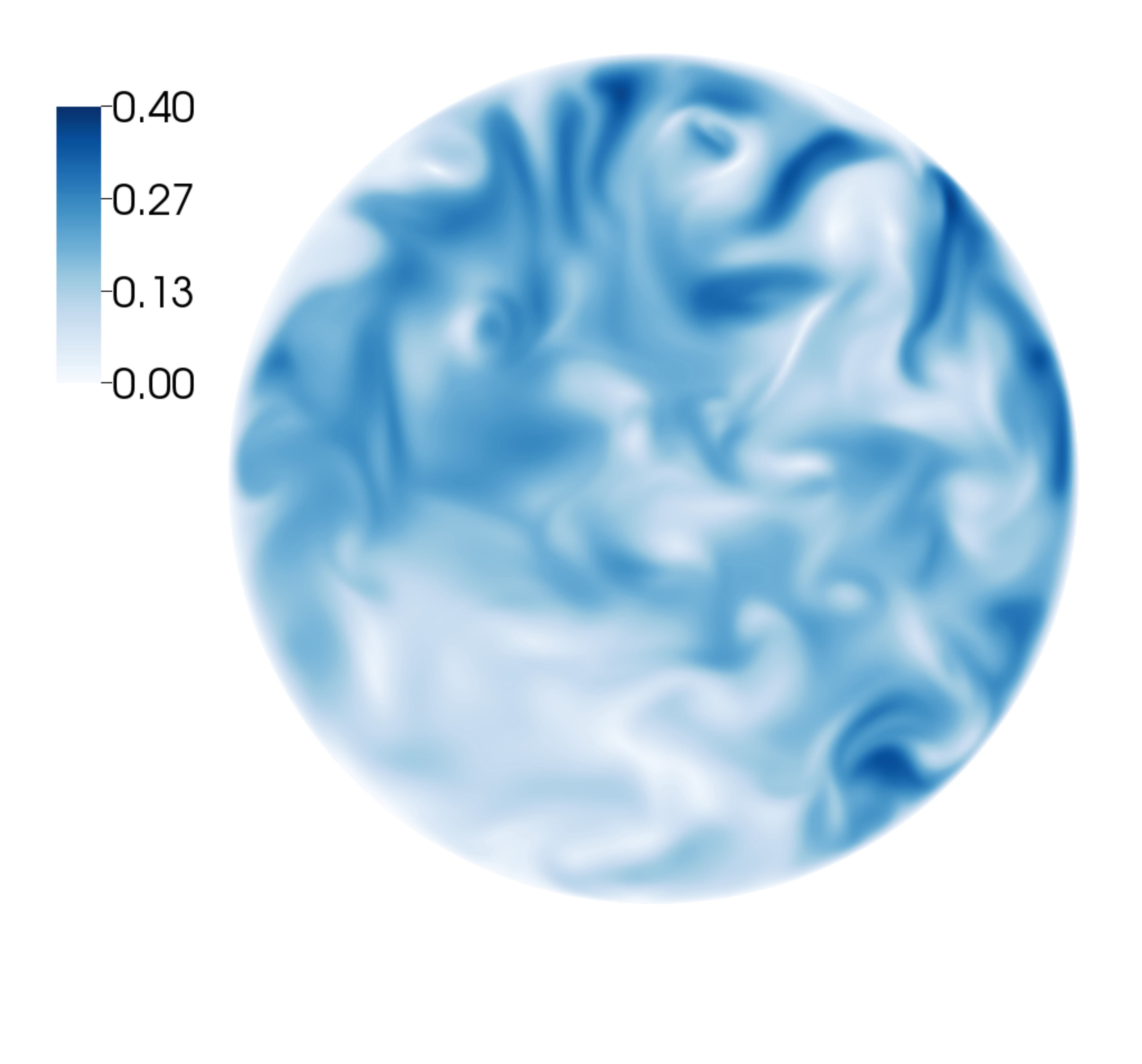}
\caption{(Colour online) Snapshots of horizontal cross sections at mid-plane of temperature field (left) and velocity magnitude (right) for $Ra=10^8$ and 
$Pr=0.021$ (top row) and $Pr=0.7$ (bottom row). The data for $Ra=10^8$ and $Pr=0.7$ have been taken from \cite{Scheel2014}.}
\label{tmgplots1}
\end{figure}
\begin{figure}
\centering
\includegraphics[width=0.5\textwidth]{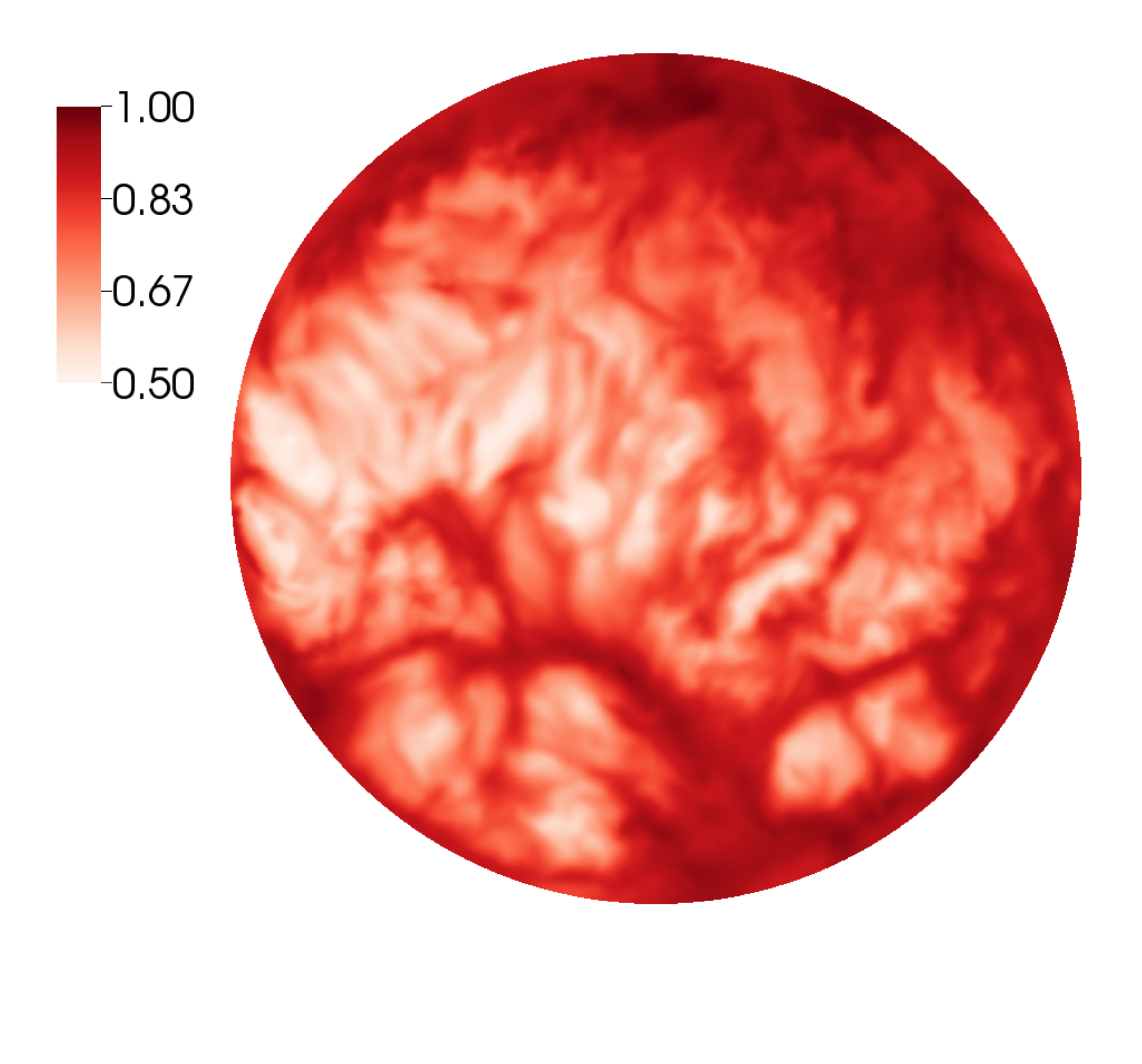}\includegraphics[width=0.5\textwidth]{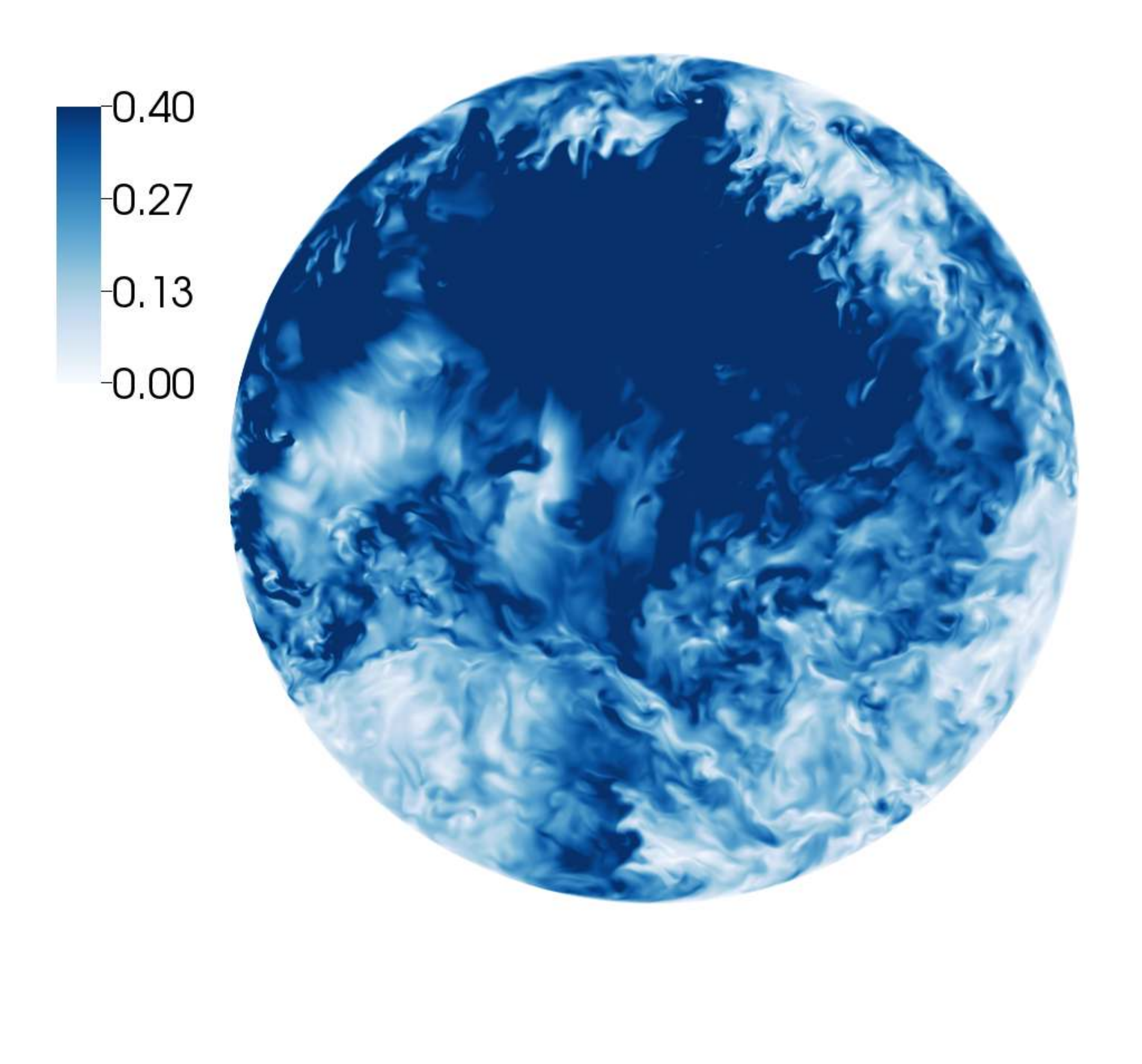}
\includegraphics[width=0.5\textwidth]{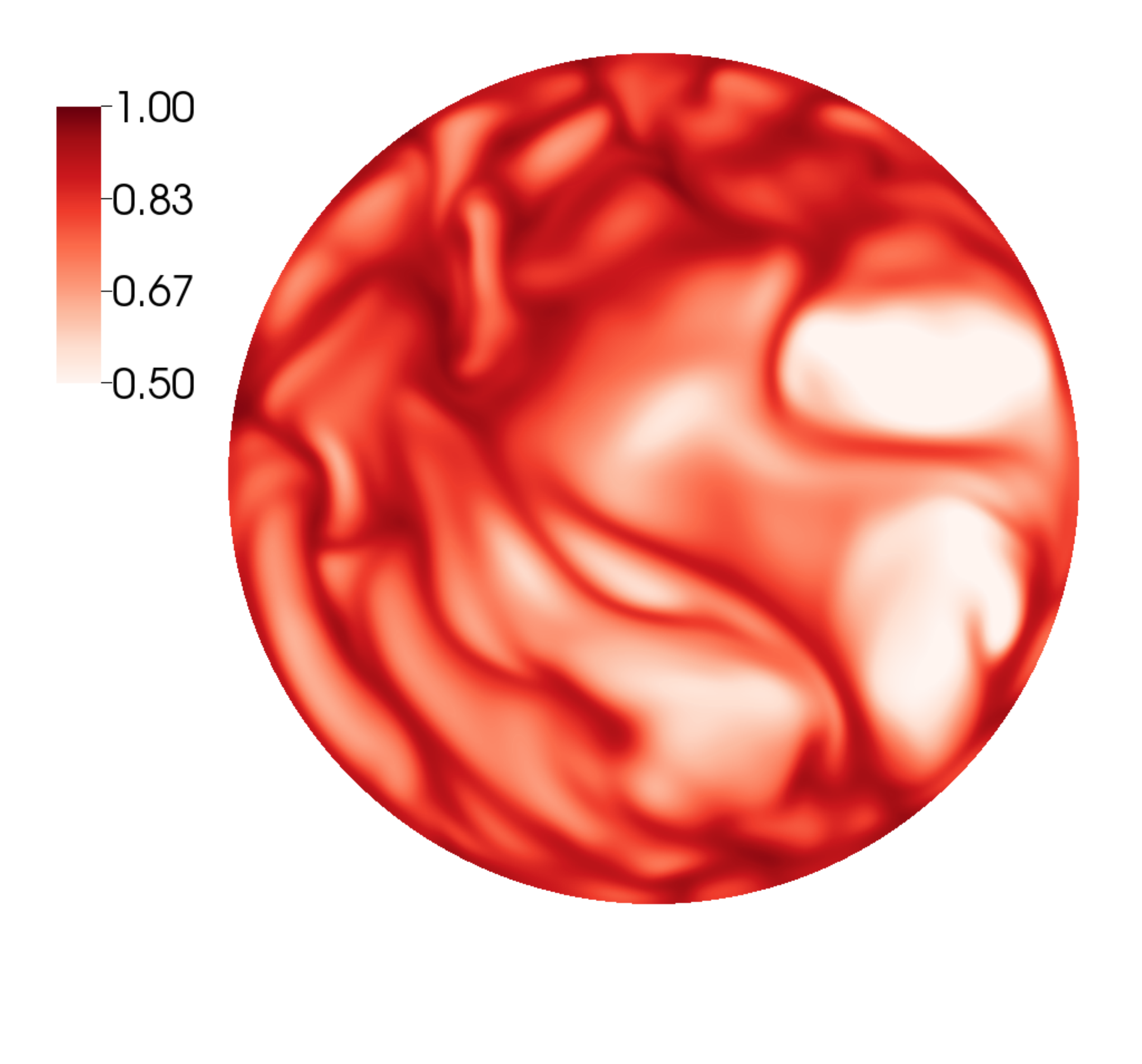}\includegraphics[width=0.5\textwidth]{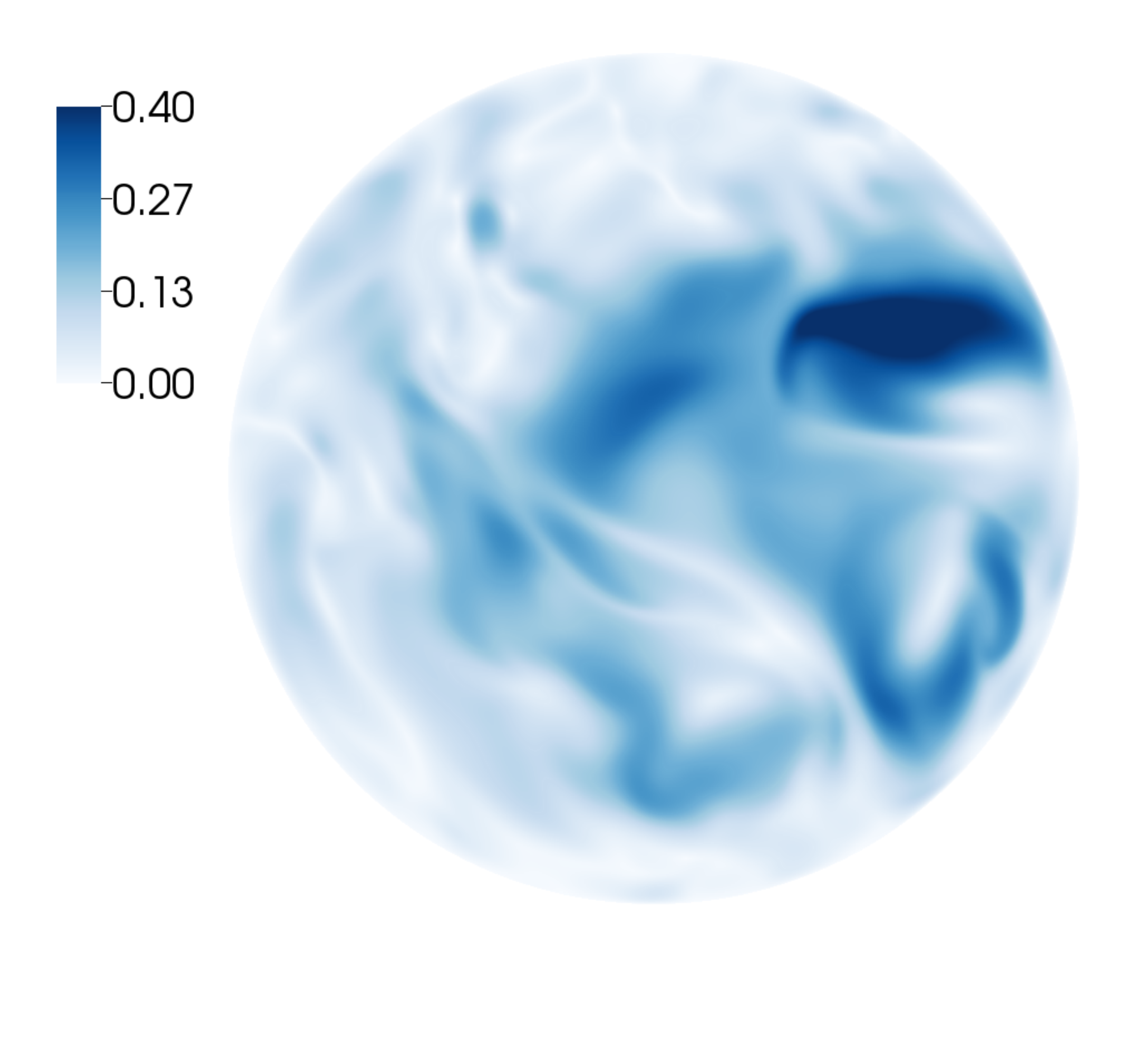}
\caption{(Colour online) Snapshots of horizontal cross sections at $\delta_T/2$ of temperature field (left) and velocity magnitude (right) for $Ra=10^8$ 
and $Pr=0.021$ (top row) and $Pr=0.7$ (bottom row). The data for $Ra=10^8$ and $Pr=0.7$ have been taken from \cite{Scheel2014}.}
\label{tmgplots2}
\end{figure}

A comparison of the instantaneous data for the same Rayleigh number of $Ra= 10^8$  and $Pr=0.021$ (top row) versus $Pr=0.7$ (bottom row) is 
shown in figures~\ref{tmgplots1} and \ref{tmgplots2}. Color density plots for both the temperature (left column) and velocity magnitude (right column) 
are shown for a representative snapshot. A horizontal cut through the cell ($z=1/2$) and inside the thermal boundary layer ($z=\delta_T/2$) is shown in 
figures~\ref{tmgplots1} and \ref{tmgplots2}, respectively.  We see in both cases that the temperature is much more diffuse in the lower $Pr$ case, but 
the velocity  is both higher in magnitude and exhibits finer structure and larger fluctuations. This combination of a diffuse temperature field coupled 
with a stronger fluctuating velocity field give rise to rather different dynamics.

\section{Global turbulent heat and momentum transfer}
\begin{figure}
\centering
\includegraphics[width=0.75\textwidth]{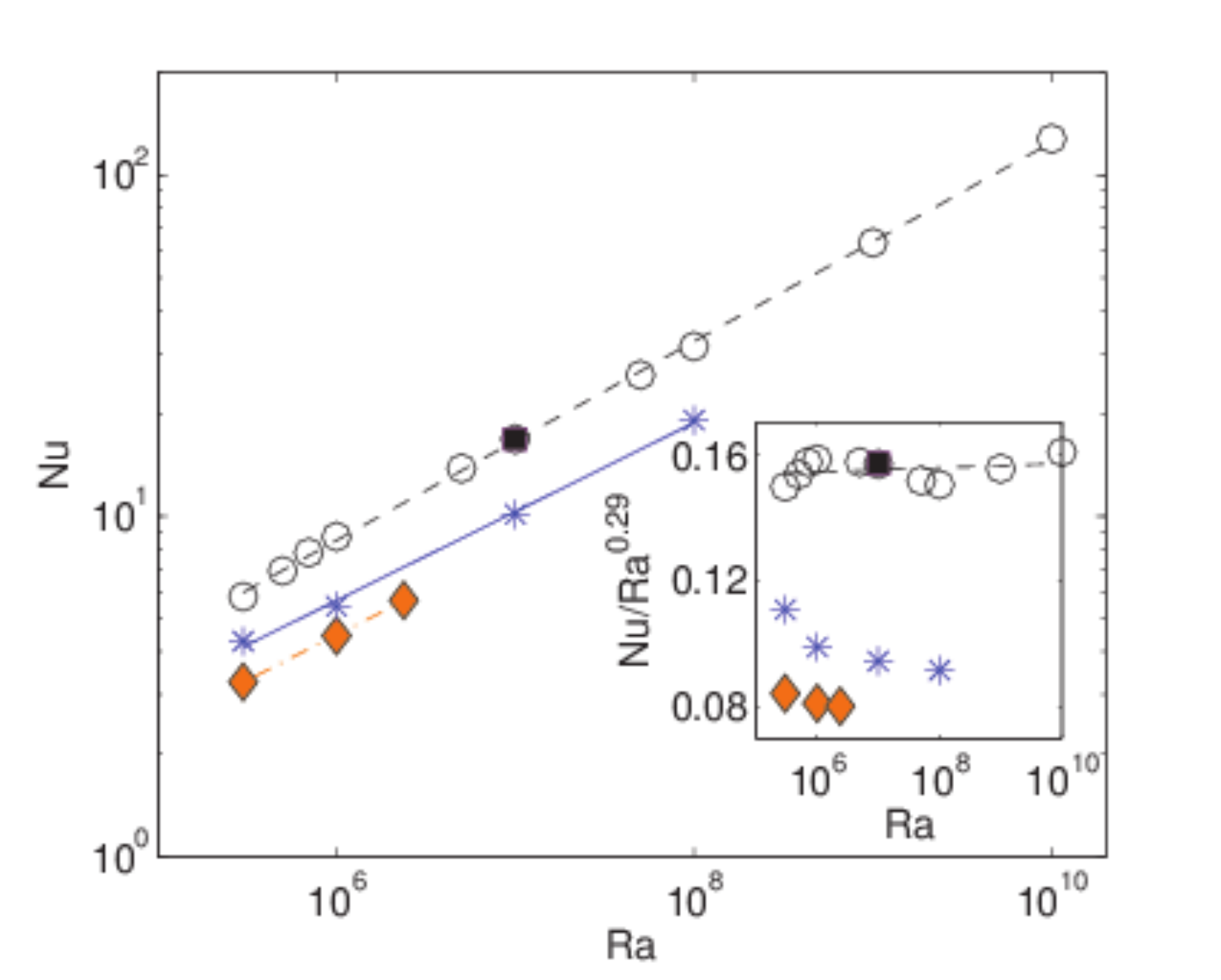}
\caption{(Colour online) Nusselt number versus Rayleigh number for $Pr=0.005$ (diamonds), $Pr=0.021$ (stars), $Pr=0.7$ (open circles), and 
$Pr=6$ (filled square). The lines are fits to the data and are given in table \ref{nurescal}. The inset replots all data compensated by $Ra^{0.29}$.}
\label{nuvsra}
\end{figure}

In figure \ref{nuvsra} we plot the Nusselt number as a function of Rayleigh number for all three Prandtl numbers (circles for $Pr=0.7$, stars for 
$Pr=0.021$ and diamonds for $Pr=0.005$).  We also added the data for $Pr=6$ which we also have for $Ra=1\times 10^7$ (see \cite{Scheel2013}). 
As expected there is a decrease in heat transport for a given Rayleigh number as Prandtl number decreases. We find the scaling exponent decreases from $0.29\pm 0.01$ to $0.26\pm 0.01$ to $0.265\pm 0.01$ 
for $Pr=0.7, 0.021, 0.005$, respectively (see table~\ref{nurescal}). It should be stressed that the present data base for the 
low--Prandtl--number simulations is still rather small.

\begin{table}
\begin{center}
\begin{tabular}{ c  r  r   r }
Quantity &  $Pr=0.7$ & $Pr=0.021$ & $Pr=0.005$   \\ \\
$Nu$ & $(0.15\pm0.01)Ra^{0.29\pm 0.01}$ & $(0.15\pm 0.04)Ra^{0.26\pm 0.01}$ & $(0.11\pm 0.02) Ra^{0.265 \pm 0.01}$ \\
$Re$ &$(0.24\pm0.01)Ra^{0.49\pm 0.01}$  & $(7.4\pm 0.6)Ra^{0.44\pm 0.01}$ &  $(9 \pm 1)Ra^{0.49\pm 0.01}$ \\
\end{tabular}
\caption{Fits to the data in figure \ref{nuvsra} (for Nusselt number $Nu$ versus Rayleigh number $Ra$) and figure \ref{revsra} 
(for Reynolds number $Re$ versus $Ra$). The uncertainty is determined by the scatter in the data and is the result of 
performing a least squares fit}.
\label{nurescal}
\end{center}
\end{table}
\begin{figure}
\centering
\includegraphics[width=0.75\textwidth]{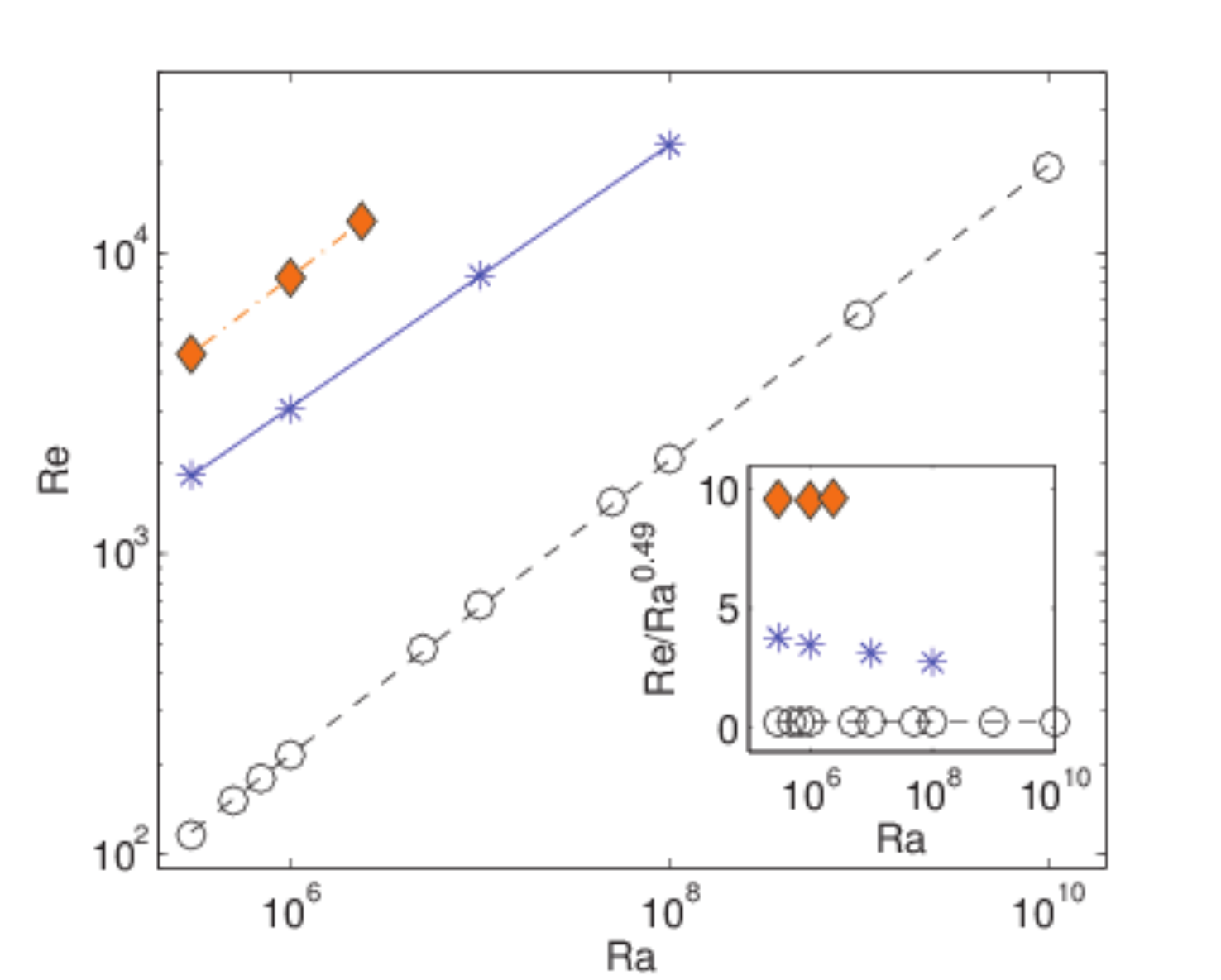}
\caption{(Colour online) Reynolds number versus Rayleigh number for  $Pr=0.005$ (diamonds), $Pr=0.021$ (stars) and 
$Pr=0.7$ (open circles). The lines are fits to the data and are given in table \ref{nurescal}. The inset replots all data 
compensated by $Ra^{0.49}$.}
\label{revsra}
\end{figure}
\begin{figure}
\centering
\includegraphics[width=0.8\textwidth]{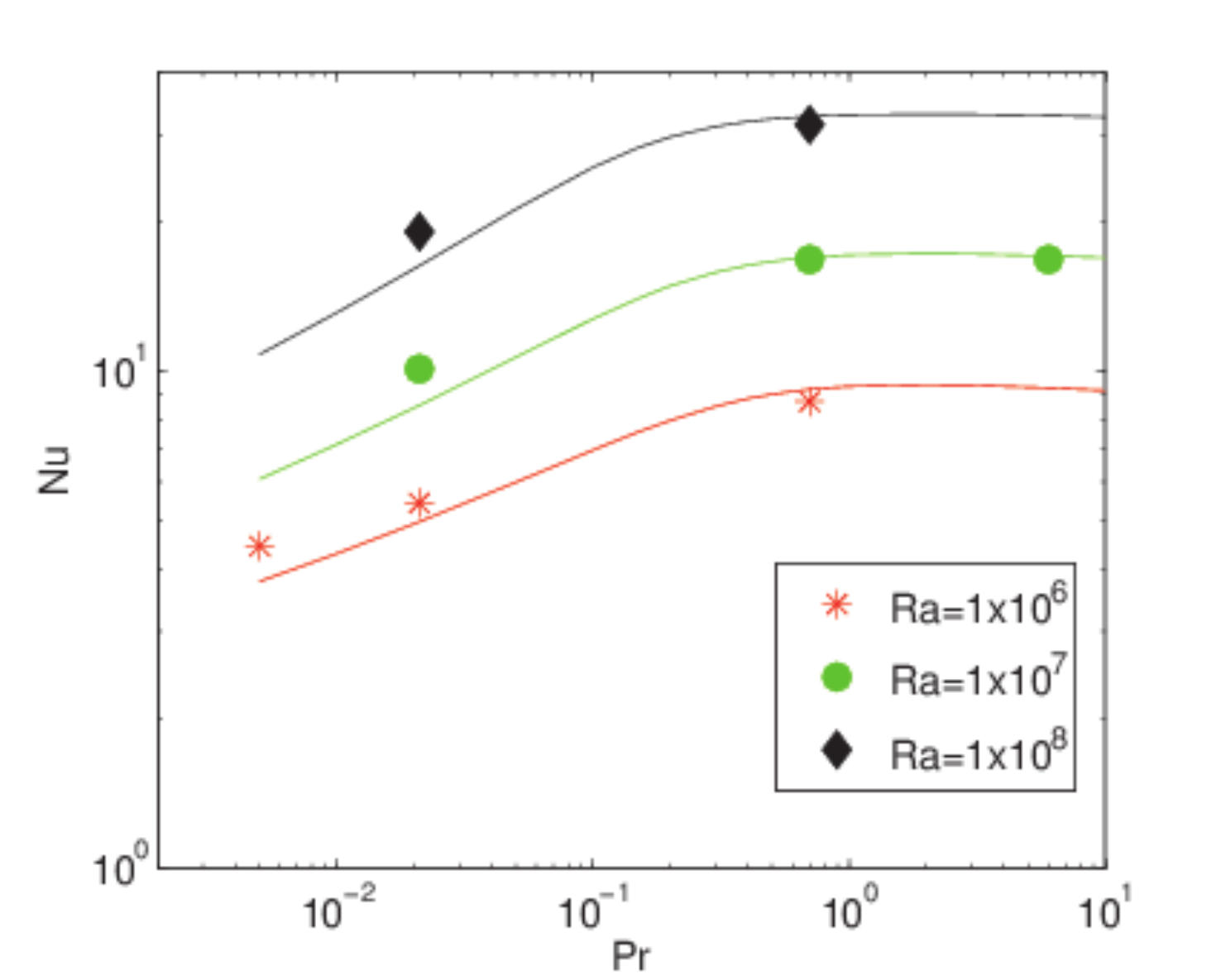}
\caption{(Colour online) Nusselt number versus Prandtl number for three different Rayleigh numbers which are indicated in the legend. 
The solid lines are the fit to the Grossmann-Lohse equations as given in \cite{Stevens2013}.}
\label{nuvspr}
\end{figure}

The scaling results are compared with previous published results in table
\ref{tab:scaling}, where the coefficients are given by
\begin{equation}
Nu = \alpha Ra^{\beta}. 
\label{compcoeff}
\end{equation}
For $Pr=0.021$, our exponent $\beta$ and prefactor $\alpha$ agree within our uncertainty with experiments by \cite{Rossby1969}, \cite{Cioni1996}, \cite{Takeshita1996},  \cite{King2013}, and simulations by \cite{Kerr2000} and  \cite{Camussi1998}. However, our exponent is on the high 
end of the range and our prefactor is on the low end of the range in table \ref{tab:scaling}.

To our knowledge, we can compare our results for $Pr=0.005$  to only one other convection experiment in liquid sodium which has 
been conducted at the Forschungszentrum Karlsruhe by \cite{Horanyi1999} and shown in table \ref{tab:scaling}. Very recent 
experiments by \cite{Frick2015} have been conducted in cylindrical cells with a very low aspect ratio of $\Gamma= 0.2$. Such a geometry 
will generate a complex large-scale circulation (LSC) pattern that consists of several rolls on top of each other rather than a single roll
as for $\Gamma\sim 1$. Therefore, these experiments are not used for comparison. Our Nusselt number value at $Ra=1\times 10^6$ is $Nu=4.45$  (see table \ref{Tabpran}) which is  higher than the 
$Nu\simeq 4.0$ which is reported by \cite{Horanyi1999}. But, our data agrees well for both the exponent and prefactor, within our 
numerical error. The conclusion which we can draw from the limited amount of data sets is that the scaling exponents $\beta$ we find 
from our DNS decrease as $Pr$ decreases, and agree with experiments and simulations when the uncertainty is taken into account. However, 
our measured exponents are 5\%  larger than the 1/4 scaling   reported in some experiments (such as \cite{King2013}) and 30\% larger than the 1/5 scaling predicted theoretically by \cite{Grossmann2000}. However, the difficulty of maintaining isothermal boundary conditions for liquid metals as noted in \cite{Horanyi1999}, should be taken 
into consideration for the experiments. It was found numerically by \cite{Verzicco2008} that when constant heat flux was used as the plate boundary conditions 
instead of isothermal, the Nusselt number for the largest Rayleigh numbers was reduced.  This was attributed to the inhibition of thermal plume growth with 
constant heat flux. This could account for the smaller experimentally measured scaling exponents.

\begin{table}
\begin{center}
\begin{tabular}{ l  c  c   c  c  c }
Group &  $\ \Gamma\ $ & $Pr$ & Range of $Ra$ & $\alpha$ & $\beta$  \\ \\
Current work* & 1 & 0.021 & $3\times 10^5-10^8$ & 0.15 $\pm$ 0.04  & 0.26 $\pm$ 0.01\\ 
\cite{Rossby1969} & 2 & 0.025 & $2\times 10^4-5\times 10^5$ & 0.147 & 0.257 $\pm$ 0.004 \\ 
\cite{Cioni1996} & 1 & 0.025 & $5\times 10^6-5\times 10^8$ & 0.14 $\pm$ 0.005 & 0.26 $\pm$ 0.02 \\ 
\cite{Camussi1998}* & 1 & 0.022 & $5\times10^4-10^6$ & --- & 0.25 \\ 
\cite{Takeshita1996} & 1 & 0.025 & $10^6-5\times 10^8$ & 0.155 & 0.27 $\pm$ 0.02 \\ 
\cite{Glazier1999} & 2,1,0.5 & 0.025 & $2\times 10^5-8\times 10^{10}$ & --- & 0.29 $\pm$ 0.01 \\ 
\cite{Kerr2000}* & 4 & 0.07 & $10^4-10^7$ & --- & 0.26 \\ 
\cite{King2013} & 1 & 0.025 & $5\times 10^6-10^8$ & 0.19 & 0.249 \\ \\
Current work* & 1 & 0.005 & $3\times 10^5-2\times 10^6$ & 0.11 $\pm$ 0.02  & 0.265 $\pm$ 0.01\\
\cite{Horanyi1999} & 4.5 & 0.006 & $2\times 10^4-5\times 10^6$ & 0.115 & 0.25 \\ 
\end{tabular}
\end{center}
\caption{Comparison of scaling coefficients for Nusselt number with Rayleigh number (see equation~(\ref{compcoeff})). 
We only selected cases which overlapped with our data range. The asterisks indicate numerical simulations; the others 
are experiments.}
\label{tab:scaling} 
\end{table}

We then investigate the dependence of Reynolds number $Re$ on Rayleigh number in figure \ref{revsra} using equation (\ref{Reynolds}) to 
compute our Reynolds numbers. We find that the Reynolds number increases rather significantly as Prandtl number decreases, which is a 
reflection of the much stronger velocity field. The scaling exponent is similar, but has decreased with Prandtl number from $0.49\pm 0.01$ 
for $Pr=0.7$ to $0.44\pm 0.01$ for $Pr=0.021$, but then returns to $0.49\pm 0.01$ for $Pr=0.005$. Experimentally \cite{Takeshita1996} 
found $Re=6.24Ra^{0.46\pm 0.02}$ for $Pr=0.025$ which agrees with our exponent and prefactor. In the simulations done by \cite{Verzicco1999} 
a scaling law $Re \simeq Ra^{0.53}$ is found which does not agree with our results.

In figure \ref{nuvspr} we show the Nusselt number as a function of Prandtl number for three Rayleigh numbers. Also included as solid lines are 
fits to the scaling theory by \cite{Grossmann2001} with the updated prefactors from \cite{Stevens2013}. The fits are very good, especially for 
$Pr=0.7$ and  6.  However we do see a consistent disagreement for the lower Prandtl numbers, with the Grossmann-Lohse theory slightly 
underestimating the value for the Nusselt number. This was also seen in the simulations by \cite{VanderPoel2013}. 

\section{Turbulence statistics for constant Grashof number}
In \cite{Schumacher2015}, we emphasized that a change in the parameter variation from constant $Ra$ to constant $Gr$ sheds a different 
light on the Prandtl number dependence in turbulent convection. Studies at constant Grashof number require a simultaneous variation of both 
the Rayleigh and Prandtl numbers as seen in figure \ref{ravspr}. In this case, the momentum equation (\ref{nseq}) remains unchanged. The 
explicit Prandtl number dependence appears only in the advection--diffusion equation (\ref{pseq}). There is however an indirect Prandtl 
number dependence in the momentum equation as well. This is manifest in the buoyancy term. A more diffusive temperature field injects 
kinetic energy at a larger scale into the convection flow, as demonstrated in \cite{Schumacher2015}.     

\begin{figure}
\centering
\includegraphics[width=0.8\textwidth]{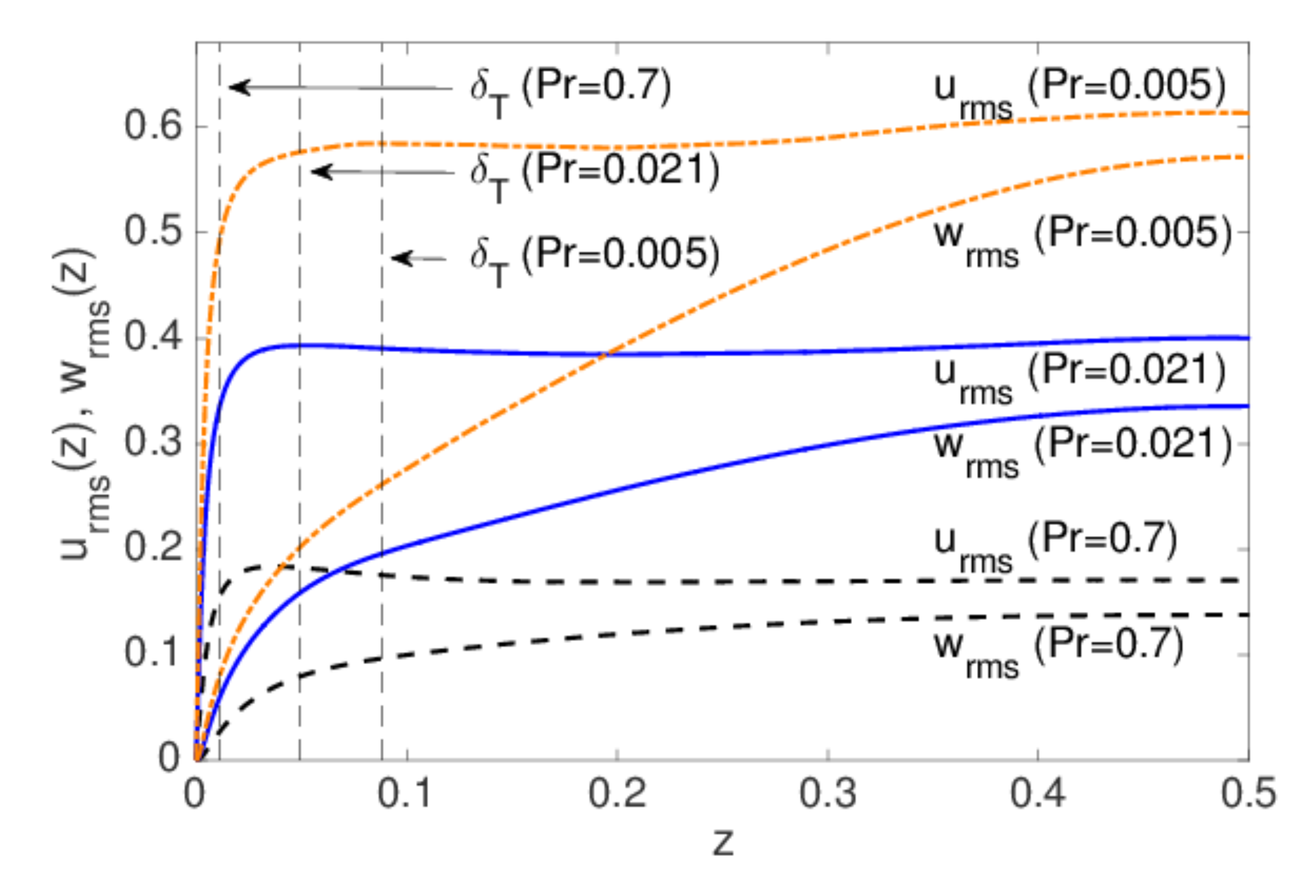}
\includegraphics[width=0.8\textwidth]{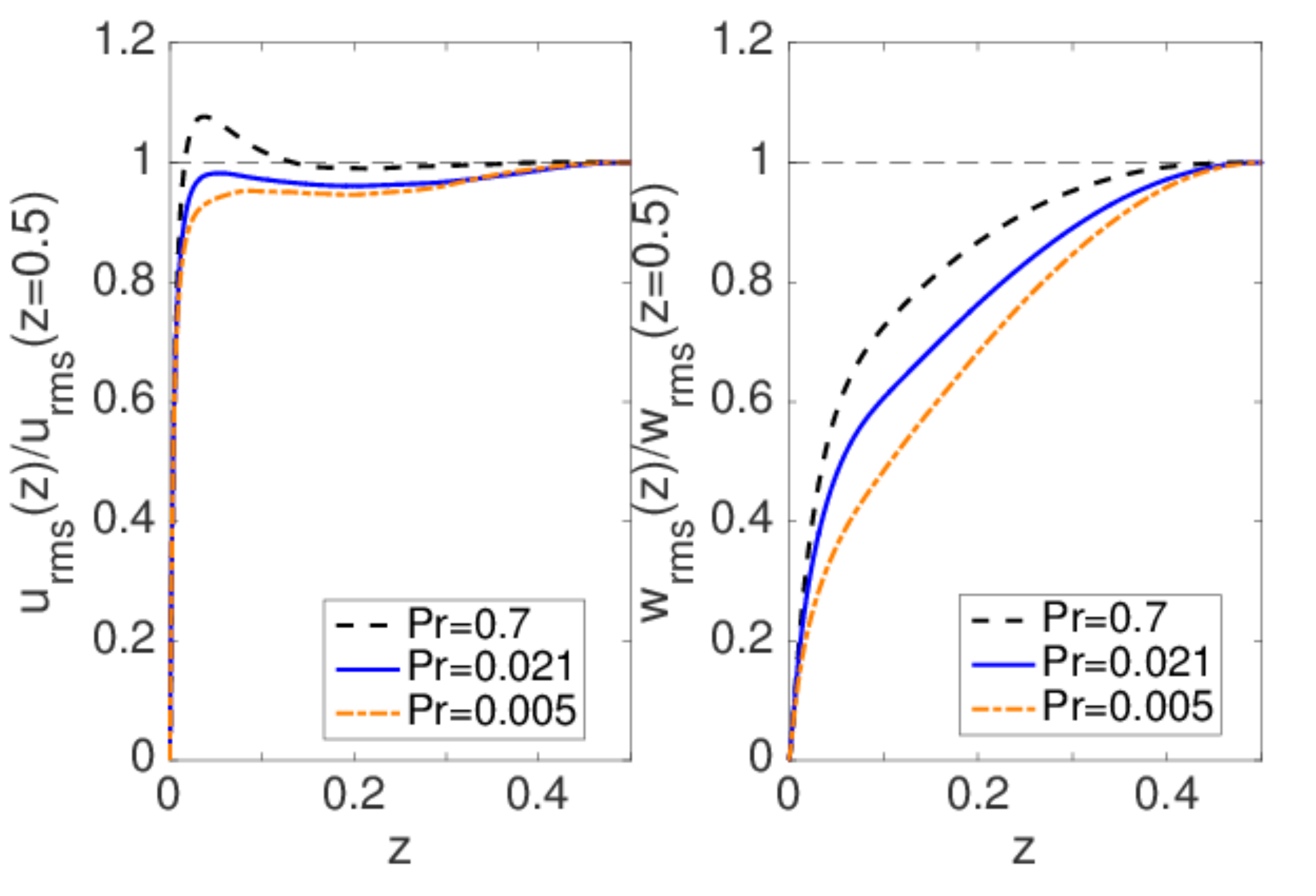}
\caption{(Colour online) Root-mean-square (rms) velocity profiles for various Prandtl numbers. Upper panel: 
The total rms velocity, $u_{rms}$, and the vertical rms velocity, $w_{rms}$, are given by equations (\ref{rms}). 
The thermal boundary layer thicknesses $\delta_T$ obtained from (\ref{TBL}) are also shown. All profiles are 
additionally averaged over the lower and upper halves of the cell. Lower left panel: Profiles of $u_{rms}$ 
collapsed with respect to their amplitude in the midplane. Lower right panel: Profiles of $w_{rms}$ collapsed
with respect to their amplitude in the midplane. All data sets are for $Gr=4.76\times 10^8$.}
\label{verrms}
\end{figure}

\begin{figure}
\centering
\includegraphics[width=0.8\textwidth]{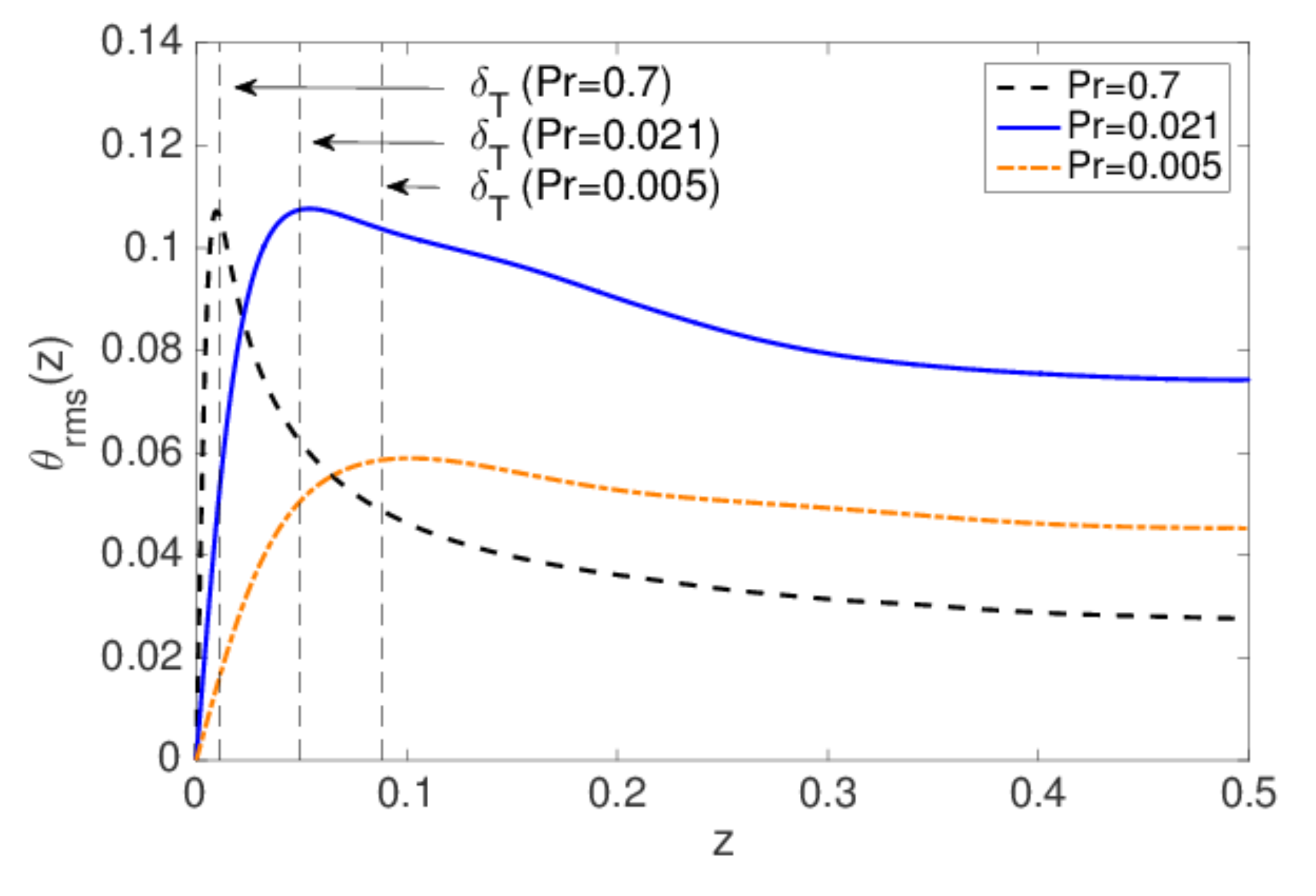}
\includegraphics[width=0.8\textwidth]{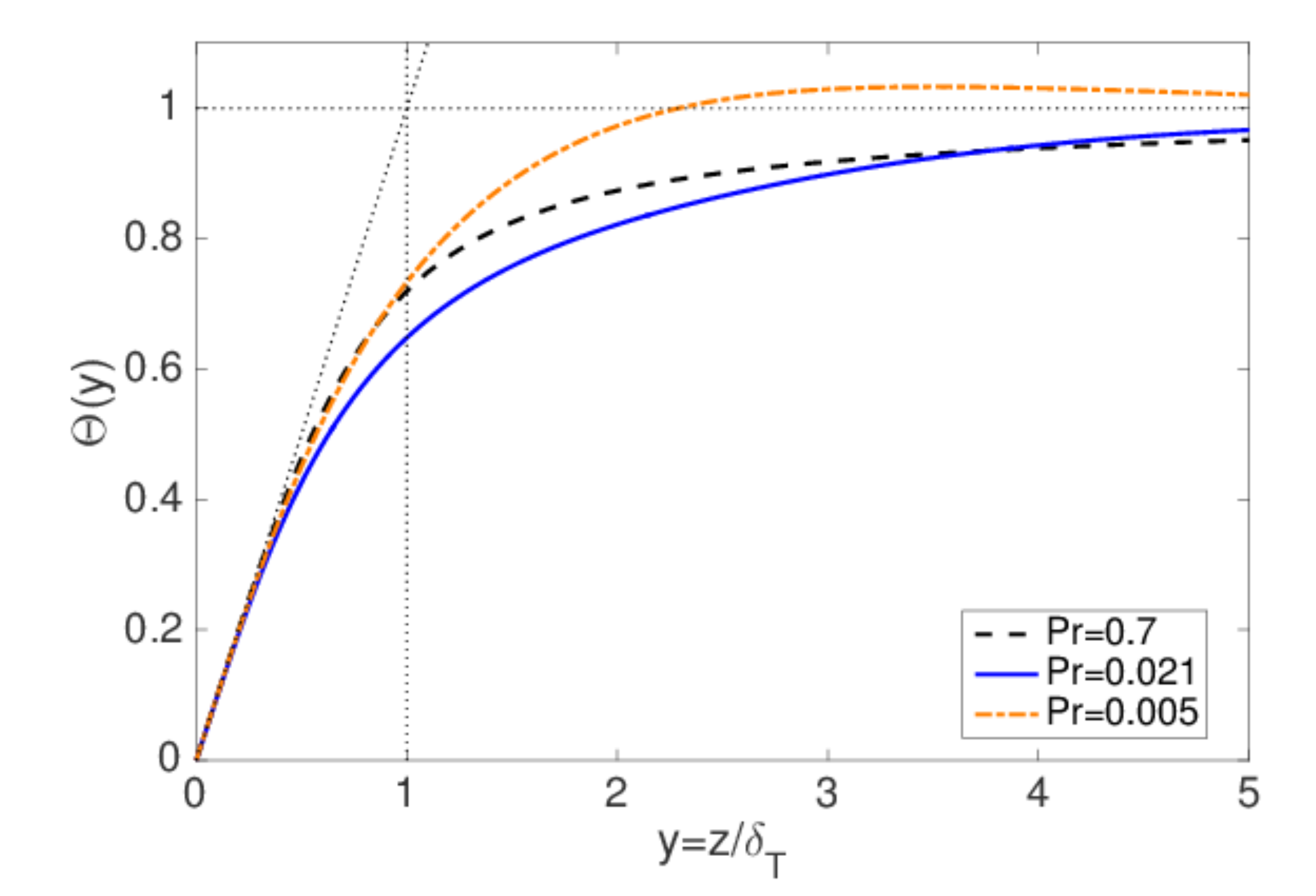}
\caption{(Colour online) Temperature field statistics for various Prandtl numbers. Top: The time averaged temperature deviation 
$\theta_{rms}$  is given by equation (\ref{rms}). The thermal boundary layer thicknesses $\delta_T$ obtained from (\ref{TBL}) are also shown. All 
profiles are additionally averaged over the lower and upper halves of the cell. Bottom: Scaled mean temperature profiles $\Theta(y)$ versus
$y=z/\delta_T$ (see equation (\ref{Tscale})).  All data sets are for $Gr=4.76\times 10^8$.}
\label{vertrms}
\end{figure}

In figures \ref{verrms} and \ref{vertrms} we compare root-mean-square (rms) profiles for different Prandtl number but fixed Grashof number
for runs 1, 4 and 8. The rms values are defined as follows:
\begin{equation}
u_{rms}(z) =\sqrt{\langle u_i^2\rangle_{A,t}}\,, \;\;\; w_{rms}(z) =\sqrt{\langle u_z^2\rangle_{A,t}}\,, \;\;\; \theta_{rms}(z) =\sqrt{\langle (T-\langle T\rangle_{A,t})^2\rangle_{A,t}}\,.
\label{rms}
\end{equation}
The upper panel of figure \ref{verrms} displays the expected increase of the velocity fluctuations, and thus of the turbulent 
kinetic energy, with decreasing Prandtl number. The vertical velocity fluctuations, $w_{rms}$ yield a significant contribution 
to $u_{rms}$ at all heights. As visible in the rescaled plot of lower right panel of figure \ref{verrms}, the profiles of $w_{rms}$ always monotonically increase  towards the cell center plane at $z=0.5$. They form an increasingly wider plateau around the 
center plane when the Prandtl number increases. Furthermore,  in the lower left panel of the same figure that for the lowest 
Prandtl number case, we detect that the maximum of the profiles of $u_{rms}$ is found at $z=0.5$. This is different from the case for $Pr=0.7$ where 
the fluctuations obey a maximum at $z\approx\delta_T$ which can be attributed to the finer thermal plumes which detach from both 
boundary layers at higher $Pr$.

The profiles of the rms of the temperature fluctuations are shown in the top panel of figure \ref{vertrms}. All three profiles 
obey a local maximum that is close to the thermal boundary layer thickness. The maximum amplitude of the root 
mean-square-temperature fluctuations is of about the same magnitude for the two larger Prandtl numbers and drops significantly 
for the smallest accessed Prandtl number of $Pr=0.005$. For the latter case, the temperature field is so diffusive such that a 
detachment of several separate thermal plumes is not observed anymore. Rather one pronounced hot upwelling on one side of 
the cell is found in combination with a pronounced cold downwelling on the diametral side. In other words, the LSC is driven by a 
single pair of plumes which causes the drop in fluctuations.

The bottom panel of figure \ref{vertrms} shows the rescaled mean temperature profiles for the same data in the vicinity of the bottom plate. Following 
\cite{Shishkina2009} we define
\begin{equation}
\label{Tscale}
\Theta(y)=2-2\langle T(z)\rangle_{A,t}\quad\text{with}\quad y=2 Nu\, z=z/\delta_T\,.
\end{equation} 
The dotted line indicates the slope of one of the profiles close to the wall. We see that the profiles for the low Prandtl numbers 
start to differ from the one for convection in air further away from the wall. The data for the lowest Prandtl number display an increase
of $\Theta$ above one which indicates a positive slope of the unscaled mean temperature profile in the bulk. This behavior is most probably 
 attributed to the small Rayleigh number that was used in the sodium case.

\begin{figure}
\centering
\includegraphics[width=0.8\textwidth]{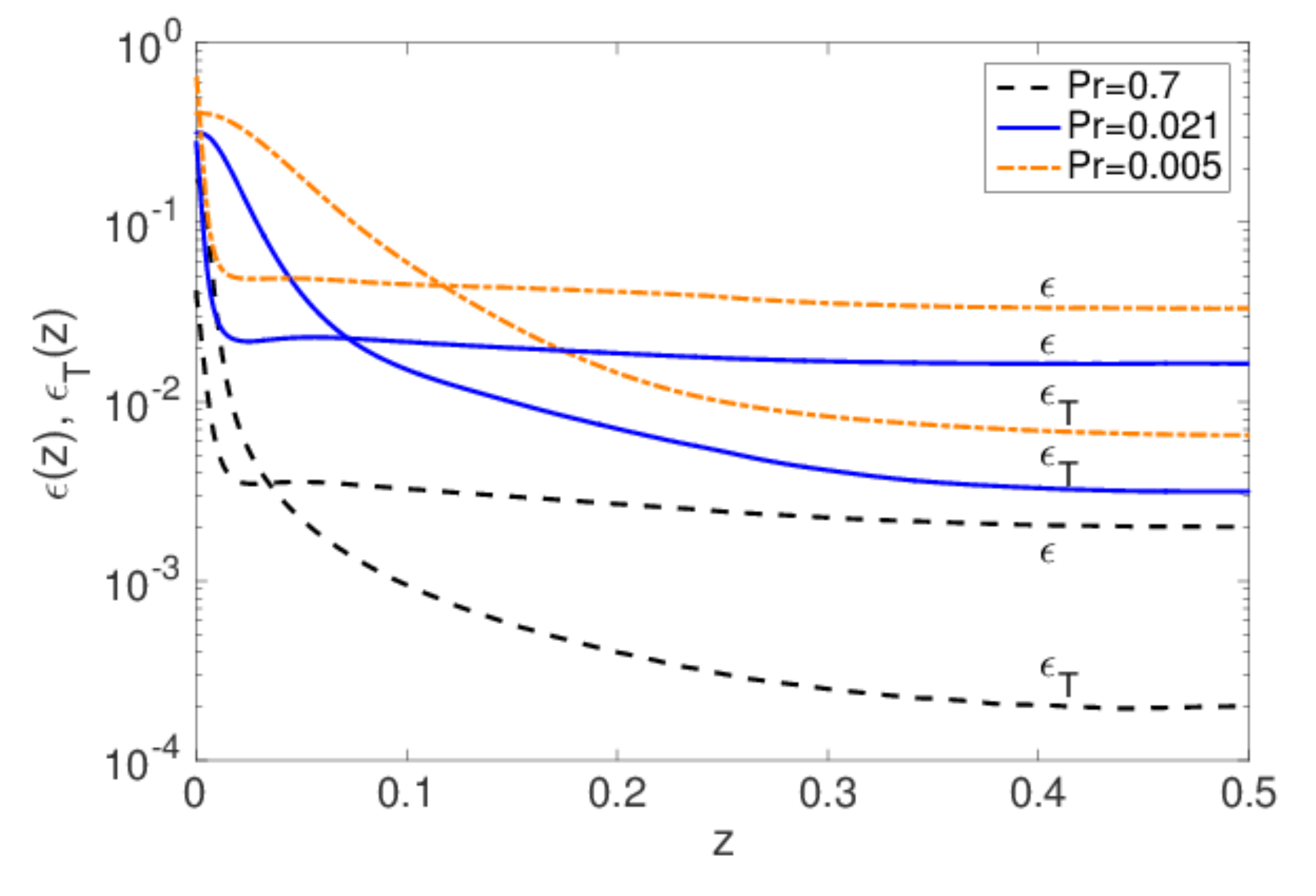}
\caption{(Colour online) Energy dissipation rate profiles for various Prandtl numbers. The kinetic energy dissipation rate $\epsilon$ 
and thermal dissipation rate $\epsilon_T$ are given by area--time averages of equations (\ref{kinetic1}) and (\ref{thermal1}), respectively. 
All profiles are additionally averaged over the lower and upper halves of the cell. All data sets are for $Gr=4.76\times 10^8$.}
\label{vereps}
\end{figure}

The vertical profiles of the plane-time averages of kinetic energy dissipation and thermal dissipation rate fields are plotted 
in figure \ref{vereps}.  They are denoted by $\varepsilon(z)=\langle\varepsilon\rangle_{A,t}$ and $\varepsilon_T(z)=
\langle\varepsilon_T\rangle_{A,t}$, respectively. The thermal dissipation rate field is given by  
\begin{equation}
{\varepsilon}_T({\bf x}, t) = \frac{1}{\sqrt{Gr}\,Pr}\left(\frac{\partial  T}{\partial  x_j}\right)^2\,,
\label{thermal1}
\end{equation}
and the kinetic energy dissipation rate field is defined as 
\begin{equation}
\varepsilon({\bf x},  t) =\frac{1}{2\sqrt{Gr}}\left(\frac{\partial  u_i}{\partial  x_j}+\frac{\partial  u_j}{\partial  x_i}\right)^2\,.
\label{kinetic1}
\end{equation}
The profiles imply that both dissipation rates increase in magnitude as the Prandtl number decreases. While the thermal dissipation 
rate grows as a result of the enhanced thermal diffusivity, the kinetic energy dissipation is increased as a result of the enhanced 
small-scale intermittency. The latter result has to be a consequence of enhanced local velocity derivatives since 
the prefactor in (\ref{kinetic1}) is unchanged. The ratios $Q_T=\varepsilon_T(z=0)/\varepsilon_T(z=0.5)$ and  
$Q_v=\varepsilon(z=0)/\varepsilon(z=0.5)$  give $Q_T\simeq 840, 100, 60$ and  $Q_v\simeq19, 16, 19$, respectively, for 
Prandtl numbers $Pr=0.7, 0.021, 0.005$. While the ratio of kinetic energy dissipation between boundary layer and bulk remains more or 
less unchanged, the dominance of thermal dissipation in the boundary layers for large Prandtl numbers is significantly diminished.

\begin{figure}
\centering
\includegraphics[width=0.8\textwidth]{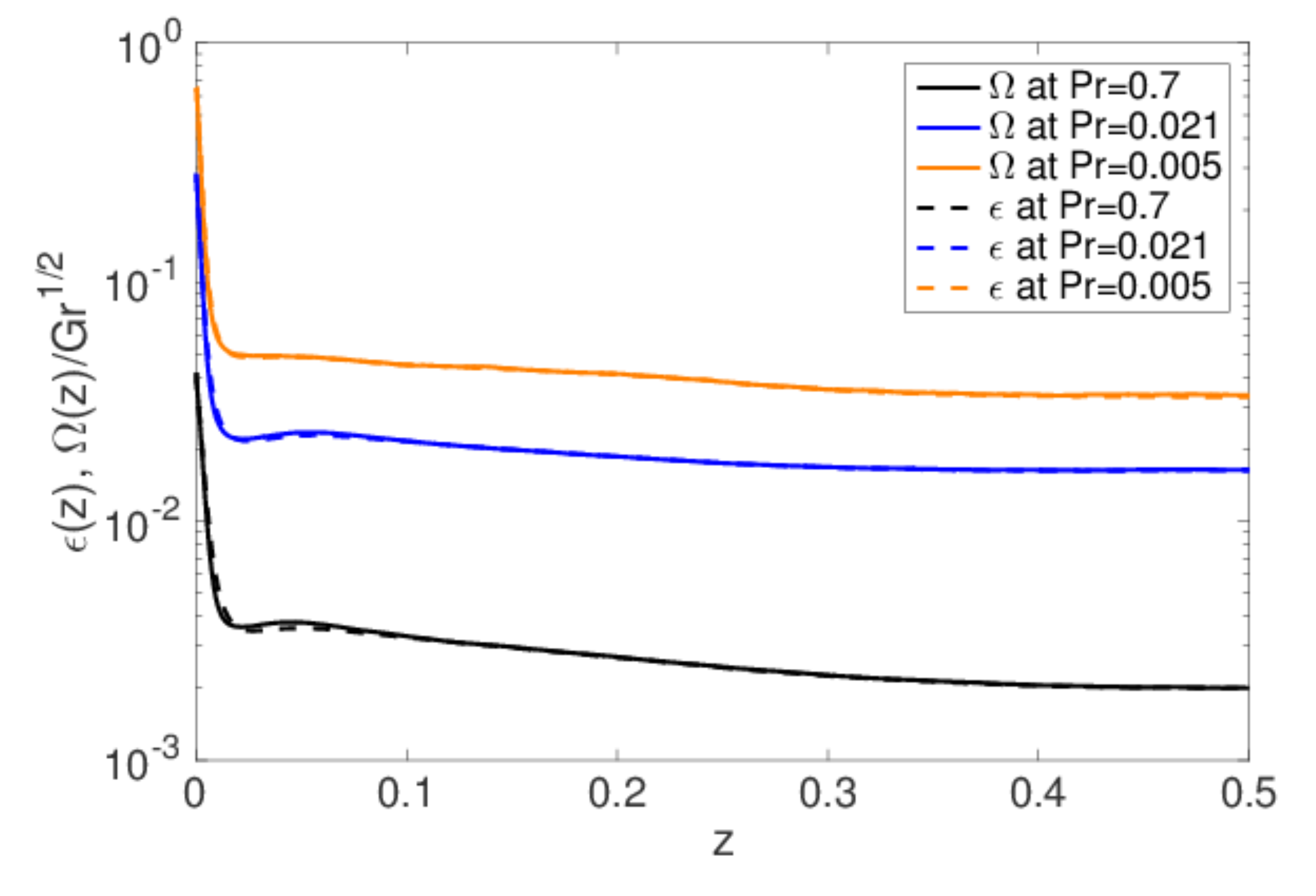}
\caption{(Colour online) Mean kinetic energy dissipation rate and rescaled local enstrophy profiles for various Prandtl numbers. All 
profiles are additionally averaged over the lower and upper halves of the cell. All data sets are for $Gr=4.76\times 10^8$.}
\label{verom2}
\end{figure}

The local enstrophy can be defined by means of the vorticity field $\omega_i=\epsilon_{ijk} \partial_j u_k$ with $i,j,k=1,2,3$ and is given by
\begin{equation}
\Omega({\bf x},t)=\omega_i({\bf x},t)\omega_i({\bf x},t)\,.
\end{equation}
In homogeneous isotropic turbulence the ensemble averages of the mean dissipation and the local enstrophy are connected by an 
exact relation which translates in our units into
\begin{equation}
\langle\varepsilon\rangle=\frac{1}{\sqrt{Gr}}\langle \Omega\rangle\,.
\end{equation}
Figure \ref{verom2} shows that this relation is satisfied to a good approximation even in our closed cylindrical cell when we compare 
the vertical mean profiles $\epsilon(z)$ and $\Omega(z)/\sqrt{Gr}$. The dashed lines for the energy dissipation rate and the corresponding solid lines for the rescaled local 
enstrophy are found to collapse almost perfectly.
\begin{figure}
\centering
\includegraphics[width=0.49\textwidth]{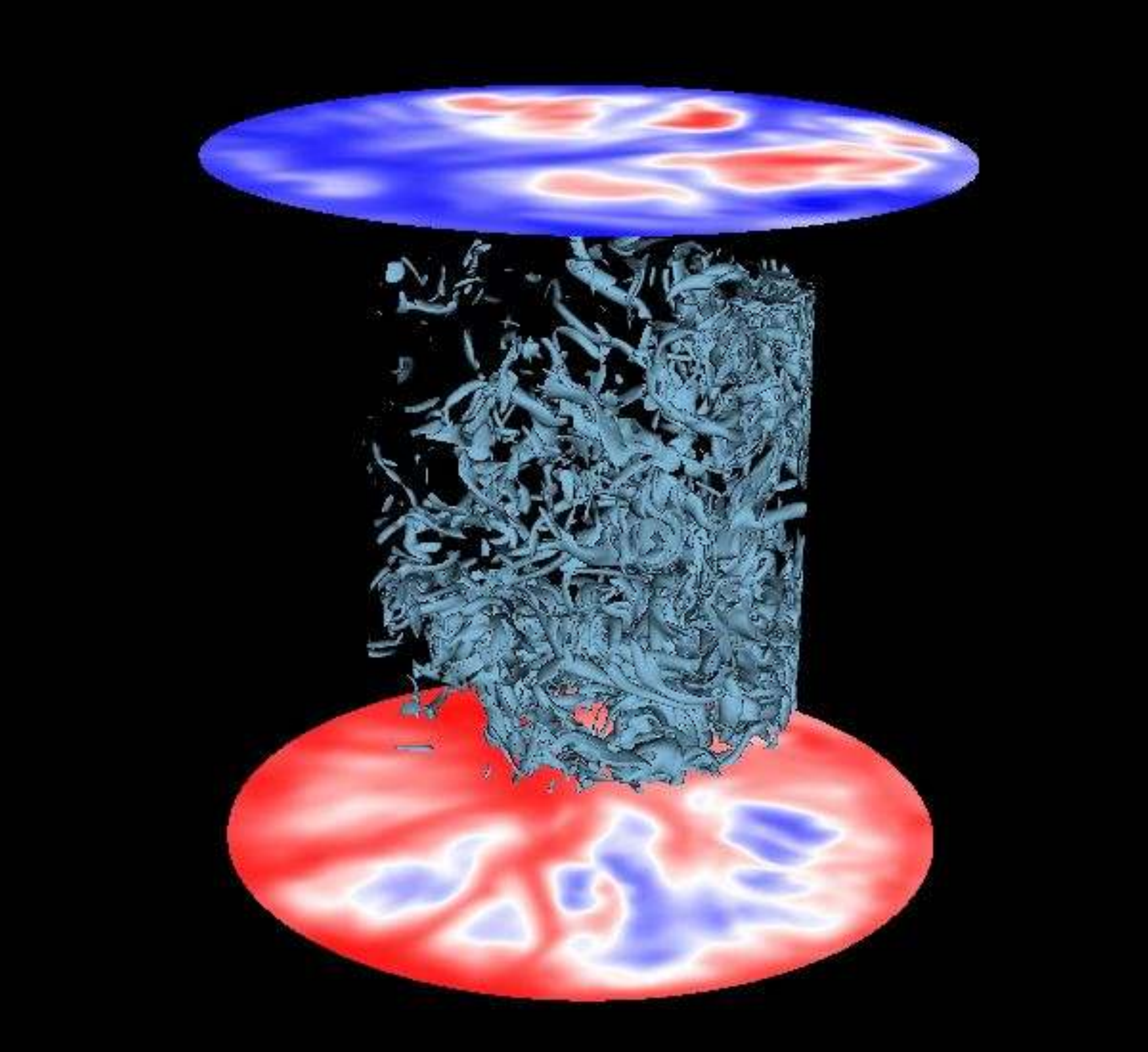} \includegraphics[width=0.49\textwidth]{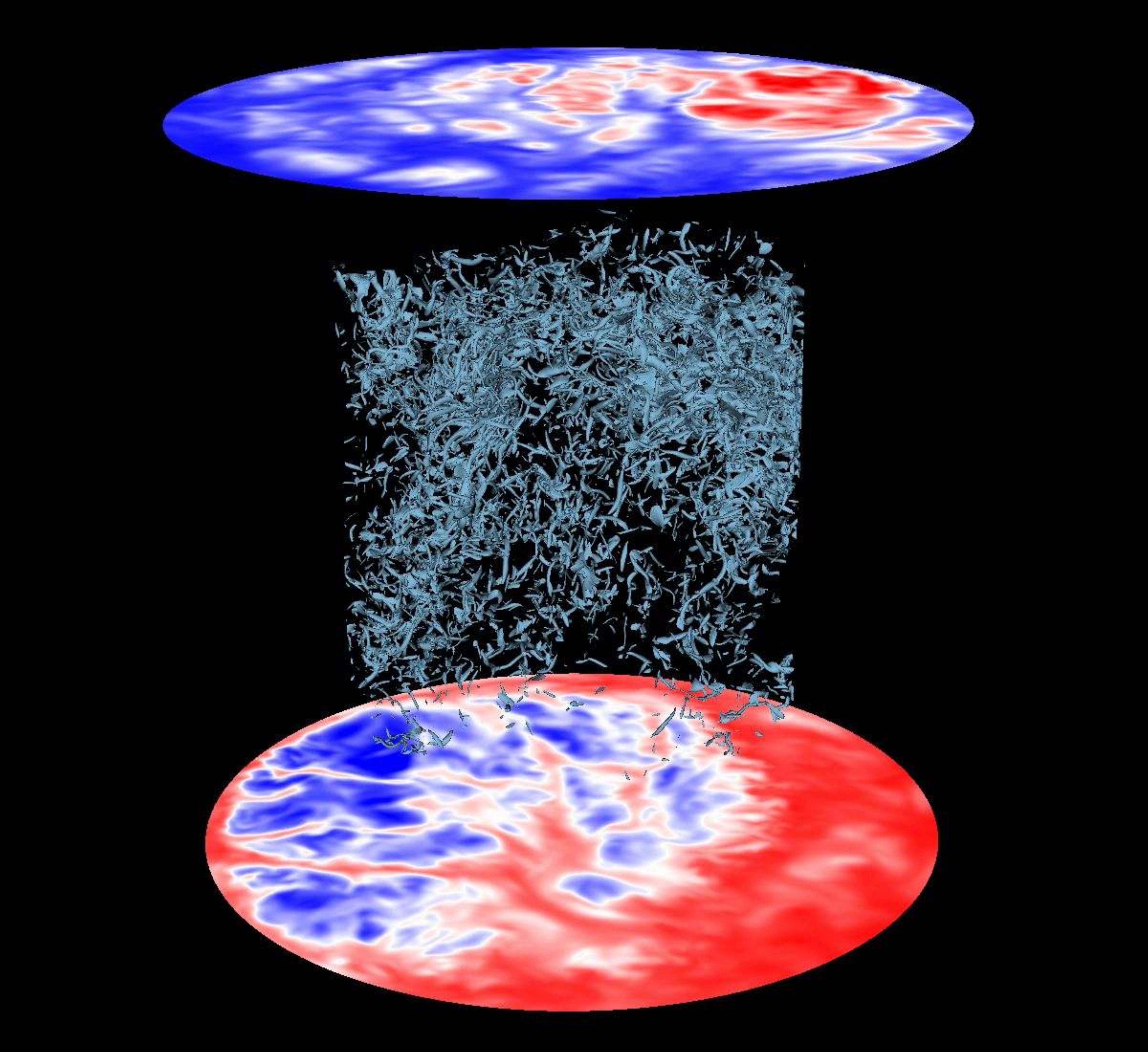}
\caption{(Colour online) Isosurfaces of constant enstrophy for $Pr=0.021$. Left panel: $Ra=10^7$, and $1\times 10^4 < \Omega 
< 1\times 10^5$. Right panel: $Ra=10^8$ and  $3.5\times 10^4 < \Omega < 5\times 10^4$. 
Also shown are the temperature field at cuts through the bottom and top boundary layers. (Movie available online.)}
\label{vort}
\end{figure}

\section{Local statistical analysis}
Isosurfaces of the local enstrophy  are plotted in figure \ref{vort} for $Pr=0.021$ for two different Rayleigh numbers, along with plots of the 
temperature field at two horizontal cross sections in the thermal boundary layers. The plot, which excludes the field at the side walls, displays 
the fine  structure of the characteristic vortex tubes, which are well-known from homogeneous isotropic turbulence (\cite{Kerr1985}). Enhanced 
small-scale intermittency was found to be in line with ever finer vortex tubes as discussed by \cite{Ishihara2009}. This is exactly what we detect 
in the center of the convection cell when comparing the two isosurface plots. In the following we want to investigate in detail if this enhanced 
intermittency in the bulk is in line with enhanced intermittency in the boundary layers, in particular in the velocity boundary layer.  

\subsection{Skin friction scaling}
\begin{figure}
  \centering
\includegraphics[width=0.5\textwidth]{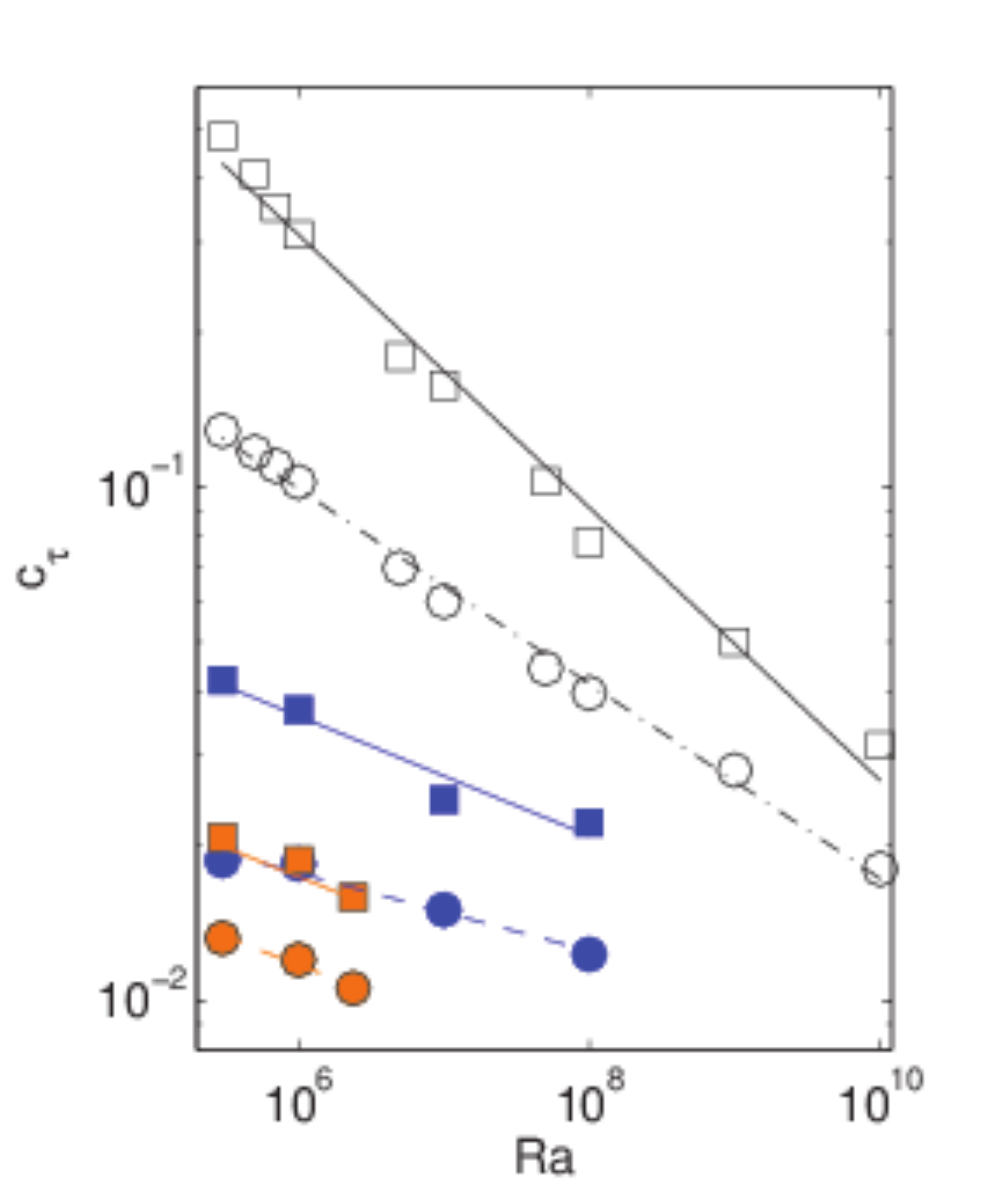}\includegraphics[width=0.5\textwidth]{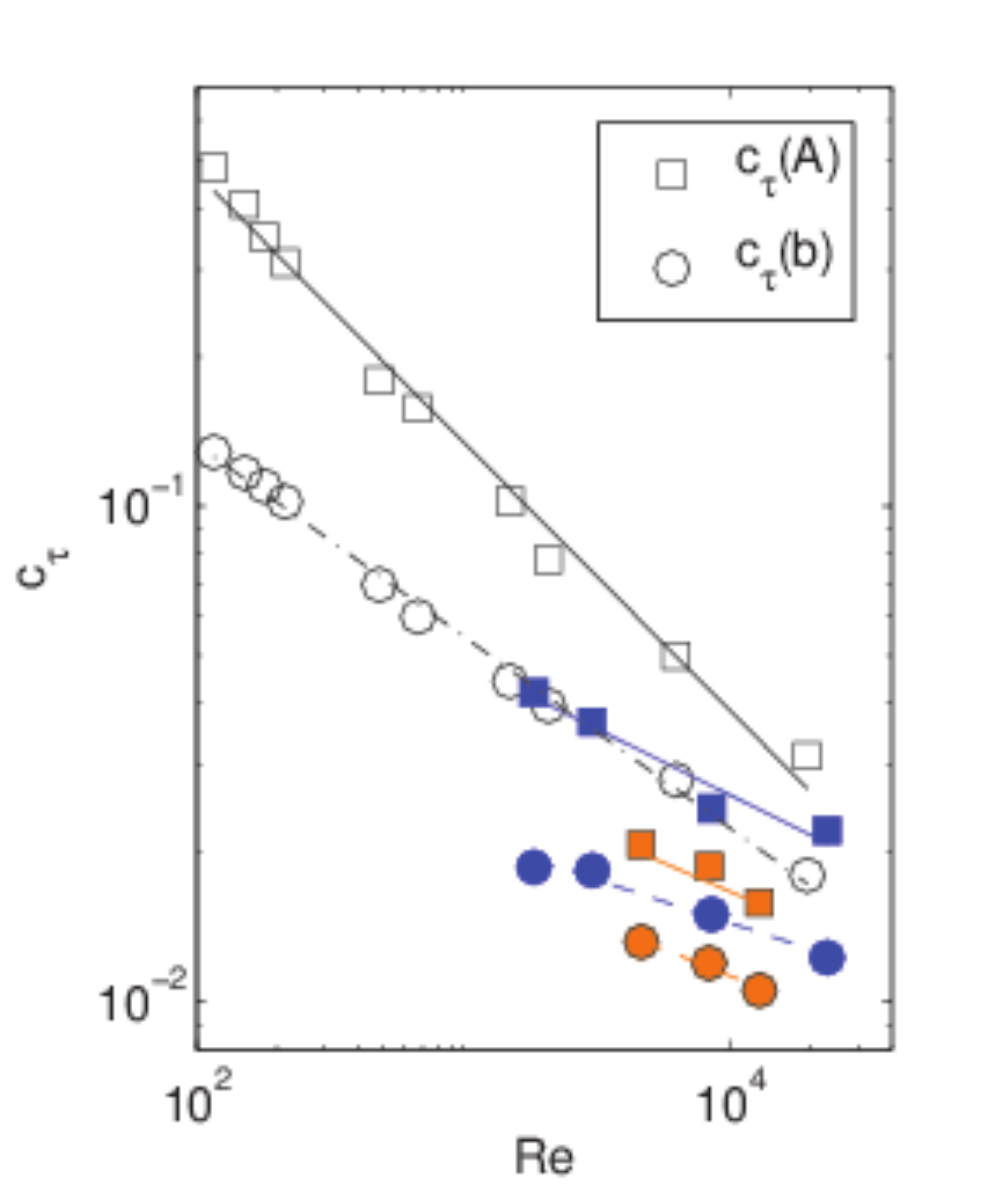}
\caption{(Colour online) Comparing friction coefficient $c_{\tau}(A)$ as defined in Eq. (\ref{ctau1}) versus Rayleigh number 
in the left panel and versus Reynolds number in the right panel, for $Pr=0.005$ (filled circles with black outline), $Pr=0.021$ (filled 
circles) and $Pr=0.7$ (open circles).  Also plotted is  $c_{\tau}(b)$ as defined in Eq. (\ref{ctau1}) versus Rayleigh number (or Reynolds number) for $Pr=0.005$ 
(filled squares with black outline), $Pr=0.021$ (filled squares) and $Pr=0.7$ (open squares). The lines are fits to the data and are 
given in table \ref{blrascal}.}
\label{tau}
\end{figure}
We start with the definition of the skin-friction coefficient (see e.g. \cite{Schlichting}) which is defined as 
\begin{equation}
c_{\tau}=\frac{\langle\hat{\tau}_w\rangle}{\hat{\rho}_0 \hat{U}^2/2}\,,
\label{cfsg}
\end{equation}
where $\langle\cdot\rangle$ is an ensemble average. The coefficient
relates the mean wall shear stress to the dynamic pressure. The hat indicates a quantity with dimension. Here, $\hat{\rho}_0$ is the  constant mass density 
of the fluid and $\hat{U}$ a characteristic velocity which will be specified further below. The following definition of the wall shear stress field at the plate is used
\begin{equation}
\label{wall}
\hat{\tau}_w(\hat{x},\hat{y},\hat{z}=0,\hat{t})= \hat{\rho}_0\hat{\nu} \sqrt{\Bigg |\frac{\partial {\hat{u}_x}}{\partial \hat{z}}\Bigg|_{\hat{z}=0}^2 + \Bigg |\frac{\partial {\hat{u}_y}}{\partial \hat{z}}\Bigg|_{\hat{z}=0}^2}\,.
\end{equation}
Following \cite{Scheel2014} this can be related to a local thickness of the velocity boundary layer by 
\begin{equation}
\label{velobl}
\hat{\lambda}_v^o(\hat{x},\hat{y},\hat{z}=0,\hat{t}) = \frac{\hat{\rho}_0\hat{\nu} \hat u_{rms,V}}{\hat{\tau}_w(\hat{x},\hat{y},\hat{z}=0,\hat{t})}\,,
\end{equation}  
excluding zero-stress events. If we assume that the $\hat{U}$ in (\ref{cfsg}) is also given by $\hat u_{rms,V}$ (see equation (\ref{Reynolds})) and 
express all variables in units of $H, U_f$ and $\Delta T$ (see section 2), we obtain the following definition for $c_{\tau}$
\begin{equation}
\label{ctau}
{c_{\tau}}= \frac{2}{Re \langle\lambda_v^o\rangle}\,,
\end{equation}
where $Re$ is defined in (\ref{Reynolds}) and the mean outer velocity boundary layer thickness is found by
\begin{equation}
\label{velBL1}
\langle \lambda^{o}_v\rangle = \int_0^{\infty}\,\lambda^{o}_v\,p(\lambda^{o}_v) \,\mbox{d}\lambda^{o}_v\,,
\end{equation}
with $p(\lambda^{o}_v)$ being the probability density function (PDF) of the outer velocity boundary layer thickness.

Rather than an ensemble average, a plane-time average is performed. Averages over the entire plate area $r\le 0.5$ will
denoted by $\langle \cdot \rangle_{A,t}$ or over the interior plate area  $r_c<0.3$ by $\langle \cdot \rangle_{b,t}$. Thus the two friction coefficients are given as 
\begin{equation}
\label{ctau1}
{c_{\tau}}(A)= \frac{2}{Re \langle\lambda_v^o\rangle_{A,t}} \quad\text{and}\quad {c_{\tau}}(b)= \frac{2}{Re \langle\lambda_v^o\rangle_{b,t}} \,,
\end{equation}
For the interior average, we selected a range of cutoff radii from $0.2 < r_c < 0.4$ and found that the results for scaling exponents and prefactors do not change significantly. However, there is a systematic trend towards smaller boundary layer thicknesses (and hence larger values of $c_{\tau}(b)$) as $r_c$ decreases. We chose $r_c=0.3$ as our cutoff radii for all plots. The uncertainty in the values in table \ref{blrascal} for interior averages is based on how the results vary as $r_c$ varies. In other words, the scaling exponent was found for $0.2 < r_c < 0.4$ and the difference between the largest and smallest scaling exponent is given as the uncertainty in scaling exponent. Likewise for the prefactor.

We have 
plotted $c_{\tau}$ as a function of Rayleigh number  and Reynolds number for $Pr=0.005, 0.021, 0.7$, and also  compare the whole area and the interior 
averages in figure~\ref{tau}. The resulting power law fits are given in table \ref{blrascal}. In all cases the skin 
friction magnitude and scaling exponent are reduced when the side-wall regions are included. For the scaling with respect to 
the Reynolds number equation (\ref{Reynolds}) is used.

Which scaling exponents can be expected for a velocity boundary layer in turbulent convection in the present Rayleigh
number range? One limiting case for the skin-friction coefficient could be that of a laminar, stationary, 
two-dimen\-sional, zero-pressure gradient flat plate boundary layer (\cite{Blasius1908}). There $c_f\sim Re_L^{-1/2}$ with 
$Re_L$ being the Reynolds number which is composed of the constant outer inflow velocity $\hat{U}_{\infty}$ and the downstream 
length $\hat{L}$ of the flat plate. The other limiting case for our present flow could be that of a fully developed turbulent flat plate boundary layer which yields a scaling law of the skin-friction coefficient which can be approximated by $c_f \sim Re_L^{-1/5}$ for 
$5\times 10^5\lesssim Re_L\lesssim 10^7$ (\cite{Coles1969}).

Clearly, these limiting cases neglect buoyancy effects which are present in the RBC flow. Furthermore, the flat plate Reynolds number $Re_L$ 
cannot be directly associated with the present $Re$. We see that for the $Pr=0.7$ case, the scaling exponent for  
$c_f$ taken for $A$ is a bit smaller than the laminar limit.  When the side-wall regions at the plate are excluded and the average is taken over $b$, 
the scaling exponent is found to be very close to the laminar limit. For the lower Prandtl numbers, the scaling with Reynolds number when the side-wall regions are included agrees within the error bars with the exponent of -0.2 for a turbulent boundary layer although the Reynolds numbers 
are by more than an order of magnitude smaller than $Re_L$. When the side walls are excluded, 
the scaling with Reynolds number is found to take values around -0.25. Not only does the bulk of the convection flow display more vigorous turbulence as $Pr$ decreases, 
but the same seems to hold for the velocity boundary layer as the global scaling of the skin-friction coefficient suggests.

\subsection{Boundary layer thickness scaling}
\begin{figure}
\centering
\includegraphics[width=0.5\textwidth]{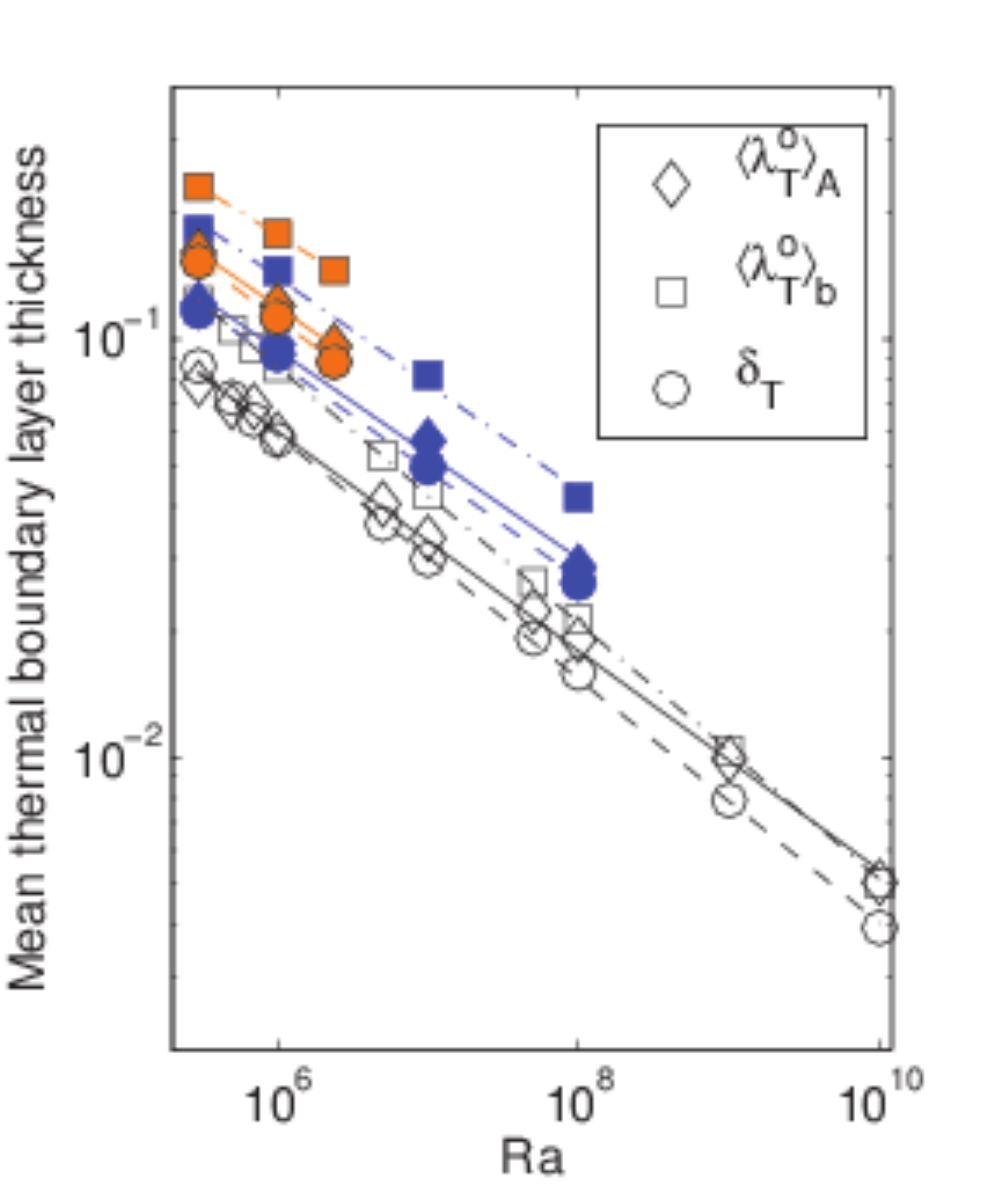}\includegraphics[width=0.5\textwidth]{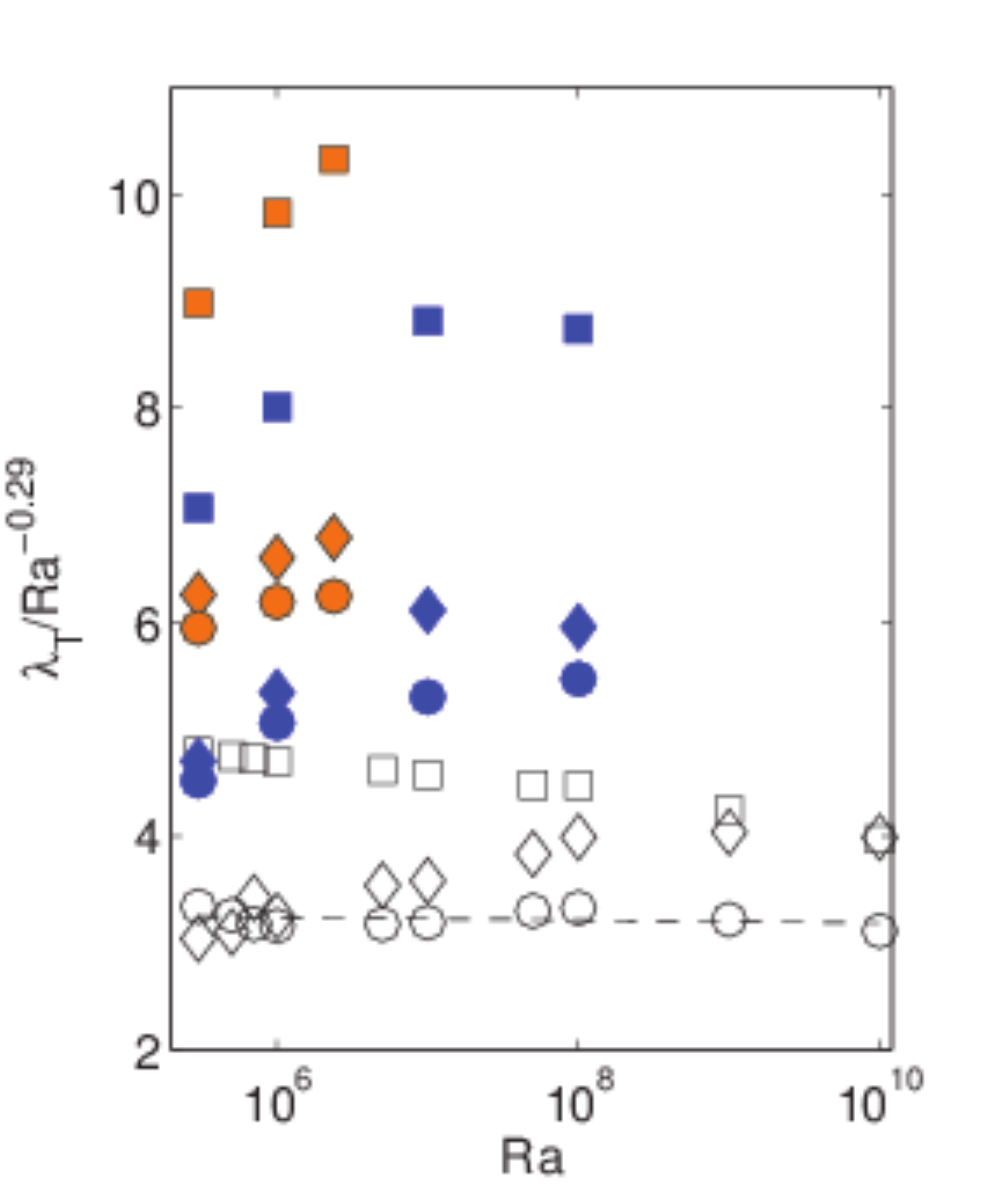}
\caption{(Colour online) Left panel: Scaling of mean thermal boundary layer thickness with Rayleigh number for different analysis methods and different Prandtl 
number. The filled data points with black outlines are for $Pr = 0.005$,  The filled data points are for $Pr = 0.021$  and the open symbols are for 
$Pr = 0.7$. The  squares for $\langle\lambda_T^o\rangle_A$ are taken over the whole plate area, the diamonds are for  $\langle\lambda_T^o\rangle_b$ 
taken over the interior area with $r<0.3$. Circles correspond to the theoretical values for $\delta_T = 1/(2 Nu)$. All data are also averaged with respect to 
time. The fits are given in table \ref{blrascal}. Right panel: the same data sets replotted where all the mean thermal boundary layer thickness data (represented by ``$\lambda_T$'')  are compensated by $Ra^{-0.29}$.}
\label{blt}
\end{figure}

We can define boundary layer thicknesses in terms of the global heat and momentum transfer (see \cite{Grossmann2000}) by 
\begin{equation}
\hat{\delta}_T=\frac{\hat{H}}{2 Nu}\,,
\label{TBL}
\end{equation}
and 
\begin{equation}
\hat{\delta}_v=\frac{a\hat{H}}{\sqrt{Re}}\,.
\label{vBL}
\end{equation}
Note that  $a$ is a free parameter (\citeauthor{Prandtl1905} \citeyear{Prandtl1905}, \citeauthor{Blasius1908}
\citeyear{Blasius1908}) which is set to $a=1/4$ in the present case following \cite{Grossmann2001}. We define the 
(dimensionless) mean thermal boundary layer thicknesses by \cite{Scheel2014} to be
\begin{equation}
\label{tempBL1}
\lambda_T^o = \frac{1}{2}\Bigg|\frac{\partial {T}}{\partial {z}}\Bigg|_{ z=0}^{-1},  \qquad \langle\lambda^{o}_T\rangle=\int_0^{\infty}\,\lambda^{o}_T\,p(\lambda^{o}_T) \,\mbox{d}\lambda^{o}_T.
\end{equation}
Figure \ref{blt} summarizes our findings for the series of low--$Pr$ runs and compares them with the data obtained by \cite{Scheel2014}. 
The scaling exponents for convection in air agree to a good approximation with those for liquid-metal convection, but with slightly decreasing 
exponent magnitude as $Pr$ decreases, which is consistent 
with the scaling for $Nu(Ra)$. The value for $\langle\lambda_T\rangle$ is always larger for the low--$Pr$ case. The 
boundary layer thickness is also larger when the side-wall regions are included. It is observed that this difference is consistently larger 
for the low--$Pr$ case, even with increasing $Ra$. The plumes that detach from the plates are coarser due to enhanced thermal diffusion 
and cause larger variations in the local boundary layer thickness.  

\begin{table}
\begin{center}
\begin{tabular}{ c  r  r   r }
  Quantity &  $Pr=0.7$ & $Pr=0.021$ & $Pr=0.005$   \\ \\
  $c_{\tau}(A)$ & $(1.4\pm 0.1)Ra^{-0.19\pm 0.01}$ & $(0.05\pm 0.01)Ra^{-0.08\pm 0.01}$ & $(0.05 \pm 0.01)Ra^{-0.11\pm 0.02}$ \\
   &  $(0.79\pm 0.05)Re^{-0.39\pm 0.01}$ &$(0.07\pm 0.01)Re^{-0.17\pm 0.02}$ &$(0.80\pm 0.03)Re^{-0.22\pm 0.03}$ \\
  $c_{\tau}(b)$ & $(12\pm 2)Ra^{-0.27\pm 0.01}$  & $(0.18\pm 0.10)Ra^{-0.12\pm 0.03}$ & $(0.08 \pm 0.03)Ra^{-0.11\pm 0.03}$   \\
     &   $(5.7 \pm 0.7)Re^{-0.54\pm 0.01}$ & $(0.31 \pm 0.1)Re^{-0.27\pm 0.02}$ & $(0.14 \pm 0.08)Re^{-0.23\pm 0.04}$ \\
$\langle\lambda_T^o\rangle_{A}$ & $(5.9 \pm 0.2)Ra^{-0.31\pm 0.01}$ & $(5 \pm 2)Ra^{-0.26\pm 0.01}$& $(3.8\pm 0.5)Ra^{-0.22 \pm 0.01}$ \\
$\langle\lambda_T^o\rangle_{b}$ & $(2.3 \pm 0.3)Ra^{-0.26\pm 0.01}$ & $(3 \pm 2)Ra^{-0.25\pm 0.02}$ & $(4.0 \pm 1 )Ra^{-0.25 \pm 0.01}$ \\
$\delta_T$ & $(3.3\pm 0.3)Ra^{-0.29\pm 0.01}$ & $(3\pm 1)Ra^{-0.26\pm 0.01}$ & $(4.4\pm 0.5)Ra^{-0.26 \pm 0.01}$\\
$\langle\lambda_v^o\rangle_{A}$ & $(5.9 \pm 0.7)Ra^{-0.30\pm 0.01}$ & $(5.4 \pm 1)Ra^{-0.36\pm 0.01}$ & $(4 \pm 1)Ra^{-0.38 \pm 0.01}$ \\
$\langle\lambda_v^o\rangle_{b}$ & $(0.55\pm 0.18)Ra^{-0.21\pm 0.01}$ & $(1.5\pm 1)Ra^{-0.32\pm 0.02}$ & $(2.5 \pm 1)Ra^{-0.38 \pm 0.03}$ \\
$\delta_v$ & $(0.50\pm 0.02)Ra^{-0.25\pm 0.01}$ & $(0.094\pm 0.003)Ra^{-0.22\pm 0.01}$ & $(0.082 \pm 0.004)Ra^{-0.25\pm 0.01}$ \\
\end{tabular}
\caption{Fits to the data in figure \ref{tau} for friction coefficient $c_{\tau}$ versus Rayleigh number $Ra$ averaged over the entire plate area ($A$) 
or over only the interior ($b$), $r<0.3$. Also given are fits to the data in figures \ref{blt} and \ref{blv} for local boundary layer thickness  versus  $Ra$ 
averaged over time and the entire plate area ($\langle\lambda_T^o\rangle_A$, and $\langle\lambda_v^o\rangle_A$) or over only the interior plate area 
($\langle\lambda_T^o\rangle_{b}$ and $\langle\lambda_v^o\rangle_{b}$). Also fit are the boundary layer thicknesses $\delta_T$ obtained from 
Nusselt number and $\delta_v$ obtained from the Reynolds number.
The uncertainty for all area averaged quantities and $\delta_T$, $\delta_v$ is determined by the scatter in the data and is the result of performing 
a least squares fit. The uncertainty in the interior averages is discussed just after (\ref{ctau1}).}
\label{blrascal}
\end{center}
\end{table}

\begin{figure}
\centering
\includegraphics[width=0.8\textwidth]{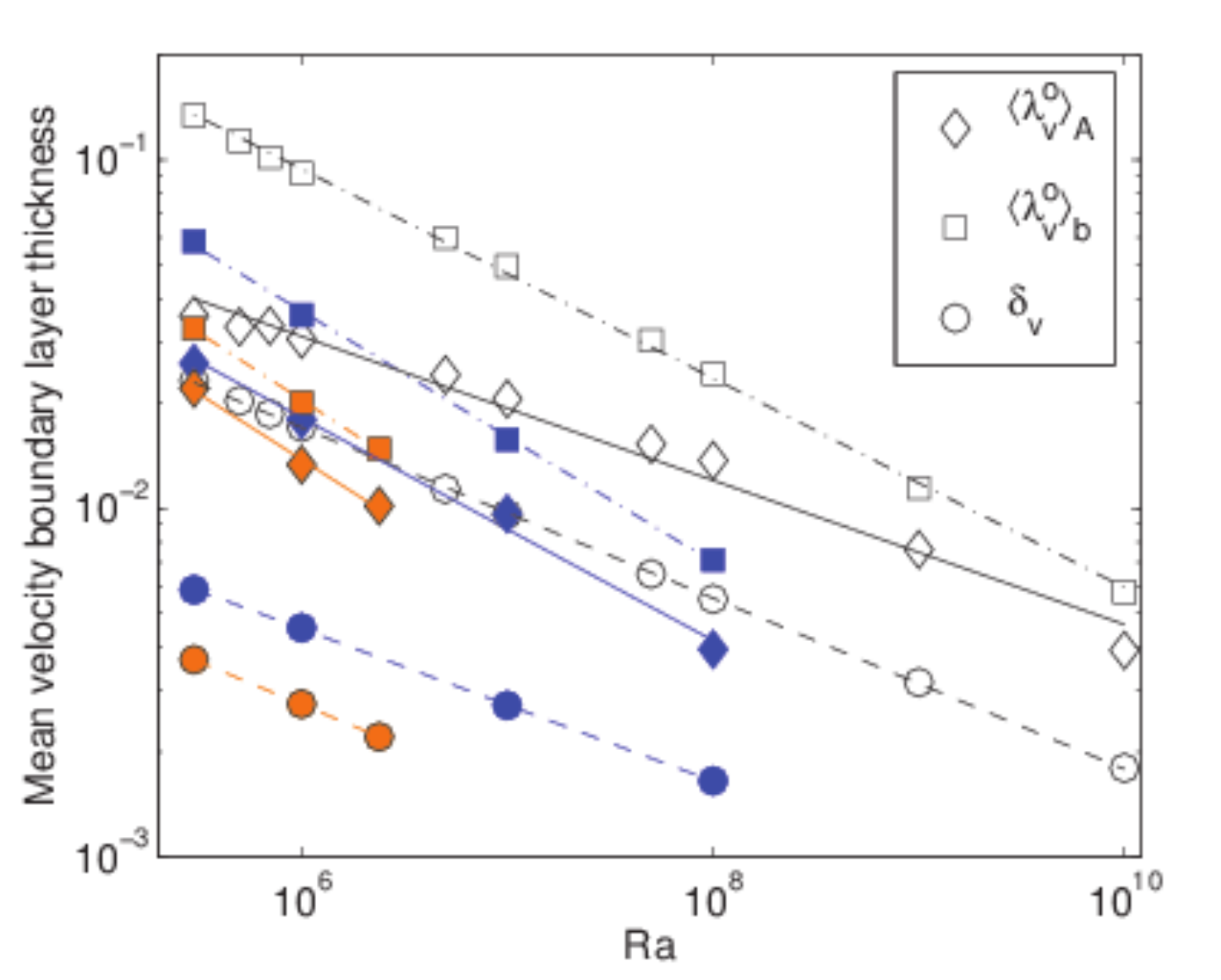}
\caption{(Colour online) Scaling of mean velocity boundary layer thickness with Rayleigh number for different analysis methods and different $Pr$. 
The filled data points with black outlines are for $Pr = 0.005$, the filled data points are for $Pr = 0.021$, and the  open symbols are for $Pr = 0.7$. The squares 
are for $\langle\lambda_v^o\rangle_{A}$ taken over the whole cell and the diamonds are for  $\langle\lambda_v^o\rangle_{b}$ taken over the interior plate area. The 
circles are the theoretical values for $\delta_v = a/\sqrt{Re}$  with the Reynolds numbers taken from table~\ref{Tabpran} and $a=1/4$. All data is averaged 
with respect to time. The fits are again listed in table \ref{blrascal}.}
\label{blv}
\end{figure}

Figure \ref{blv} summarizes our findings for $\langle \lambda_v^o\rangle$, which is defined in (\ref{velBL1}). We find that $\langle\lambda_v^o\rangle$ 
is significantly smaller for the low--$Pr$ case. We also find that the scaling exponent for  $\langle\lambda_v^o\rangle$ for the low--$Pr$ cases becomes 
larger in magnitude as $Pr$ decreases. Since  $\langle\lambda_v^o\rangle$ is determined from the inverse of the velocity gradients, this suggests that the magnitudes of these gradients 
get larger with $Ra$ at a faster rate for the lower $Pr$ case. However, the scaling exponent for $\delta_v$ as measured from the Reynolds number 
for $Pr=0.021$ case is a bit smaller in magnitude than that for convection in air ($Pr=0.7$) or liquid sodium ($Pr=0.005$), consistent with our results from figure \ref{revsra}.

\subsection{Local shear Reynolds number} 
We define the local shear Reynolds number  as 
\begin{equation}
\label{Reshloc}
Re_{sh}(x,y)=\sqrt{\frac{Ra}{Pr}} \lambda_v^{o}(x,y)u_{rms}(z=\lambda_v^{o}(x,y))\,.
\end{equation}
This is a good measure of the stability of the flow in the boundary layer. It is similar to equation (41.1) in \cite{Landau1987} but 
we pick $u_{rms}$ at $\lambda_v^o$ because the boundary layer thickness varies locally (see \cite{Scheel2014}), hence it is 
more straightforward than selecting $U$, the fluid velocity outside the space- and time-averaged boundary layer.

According to \cite{Landau1987} and \cite{Tollmien1929}, when the value of $Re_{sh} \simeq 420$ is exceeded, then the boundary 
layer is turbulent. The boundary-layer scale that was used for this estimate is the displacement thickness scale, and this applies 
to two-dimensional turbulent shear flow. We could compute the area and time averaged $Re_{sh}$ to assess when the boundary layer becomes turbulent, which was done in \cite{Scheel2014} and \cite{Wagner2012}. But, since the  locally calculated $Re_{sh}$ (\ref{Reshloc}) varies strongly and reaches values greater 
than 420  for local regions of the plate and for large enough Rayleigh number, we can also assess this spatial intermittency by finding the fractional 
area which is defined as
\begin{equation}
{\cal T}=\frac{A({Re_{sh}>Re_{sh}^c})}{A}\,,
\label{turbfrac}
\end{equation} 
where the numerator of (\ref{turbfrac}) describes the area of the plate where $Re_{sh}(x,y,t) > Re_{sh}^c$, 
and $Re_{sh}^c$ is a critical-shear-threshhold Reynolds number, i.e., 420 in the case of  \cite{Landau1987}. We  can compute ${\cal T}$ very precisely 
with the code and our area average uses the local volume element for our collocation data points. Also note that the total area $A$ is given by $\pi/4$.

\begin{figure}
\centering
\includegraphics[width=0.8\textwidth]{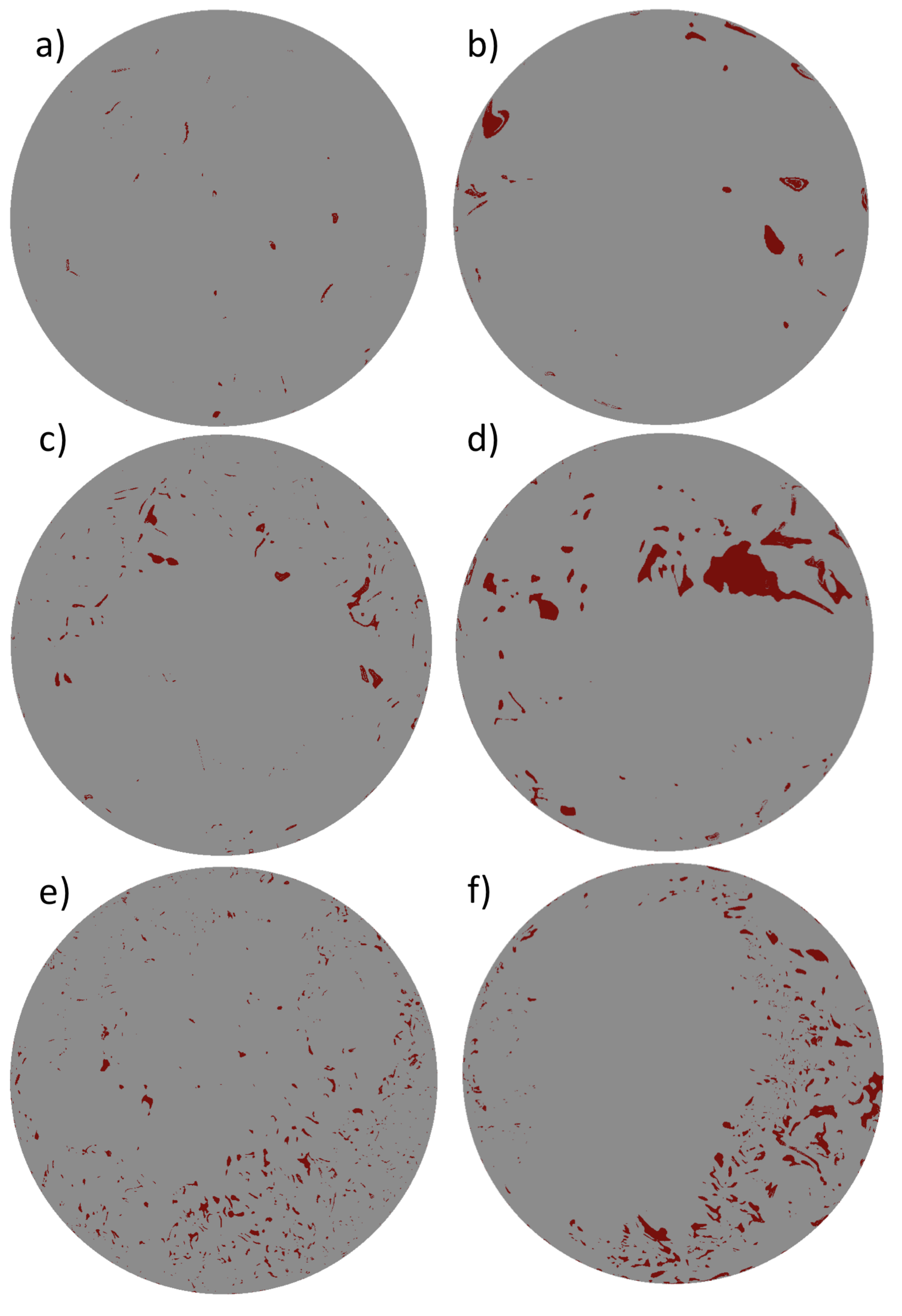}
\caption{Threshhold plots demonstrating the area of the bottom plate whose shear Reynolds number as calculated in (\ref{Reshloc}) is greater 
than 250 (red). The gray  area denotes shear Reynolds numbers less than 250. The left column is for $Pr=0.7$ and (a) $Ra=1\times 10^8$ (c) 
$Ra=1\times 10^9$, and (e) $Ra=1\times 10^{10}$ and the right column is  for $Pr=0.021$ and (b) $Ra=1\times 10^6$, (d) $Ra=1\times 10^7$.  
(f) $Ra=1\times 10^8$. }
\label{percplt}
\end{figure}

Since there is no theoretical value for the transition to turbulence in the boundary layer for Rayleigh-B\'enard convection, we take three values 
for $Re_{sh}^c: 250, 350, 450$. This  brackets a range of possible turbulent onset values (including the value of 420 from \cite{Landau1987})
and provides calculation of the uncertainty. We plot snapshots of ${\cal T}$ for various Rayleigh and Prandtl numbers in figure 
\ref{percplt} to highlight the intermittency. We chose $Re_{sh} > 250$ as our cutoff in the figures as this provides a larger region for ${\cal T}$. We see in  
figure \ref{percplt} that the structures where $Re_{sh} > 250$ are much finer for the higher $Pr$ case and increase 
more in number as $Ra$ increases.

\begin{figure}
\centering
\includegraphics[width=0.8\textwidth]{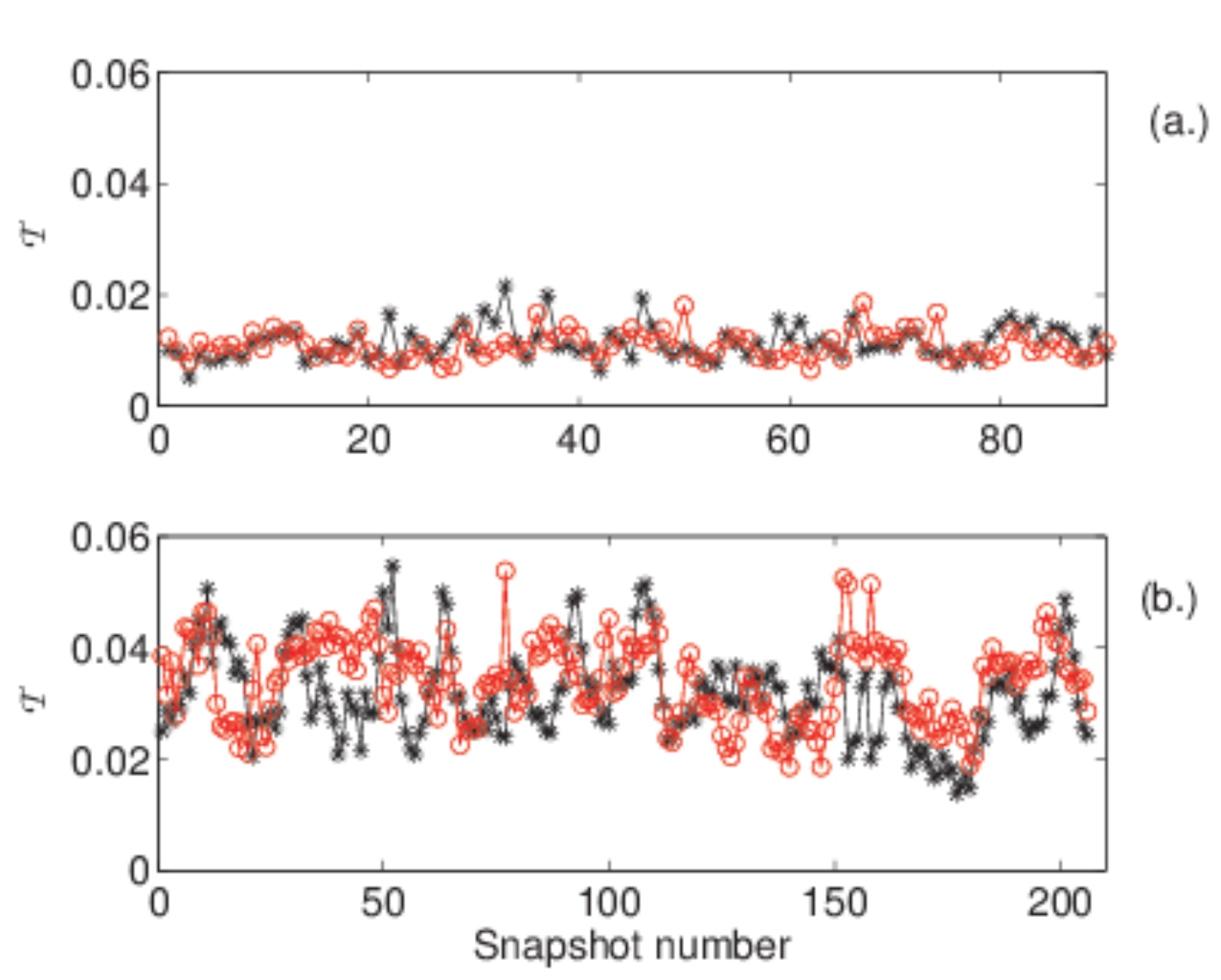}
\caption{(Colour online) Time dependence of fractional area ${\cal T}$ (defined in (\ref{turbfrac}) greater than 250 (black line=top plate,  
magenta (light gray) line=bottom plate). The fractional area is given for each statistically independent snapshot. (a) $Pr=0.7$ and $Ra=1\times 10^9$. 
(b) $Pr=0.021$ and $Ra=1\times 10^7$.}
\label{reshvt}
\end{figure}

In figure \ref{reshvt} we look at the time evolution of the fractional area, and note that it varies by about 25\%, as is 
expected for the intermittency in the boundary layer.  We also see variation between the top and bottom plate, with some anticorrelation which we attribute to the large-scale circulation. If, for example a plume is ejected from the bottom plate, causing a large value 
of ${\cal T}$ on the bottom plate, it takes some time for this plume to be swept up to the top plate by the large-scale circulation. Eventually this 
plume hits the top plate which could cause another plume to be ejected by the top plate, causing ${\cal T}$ to now be larger at the top plate. 
However, the anticorrelation is not perfect and the situation is more complex, since plume ejection occurs intermittently.

\begin{figure}
\centering
\includegraphics[width=0.8\textwidth]{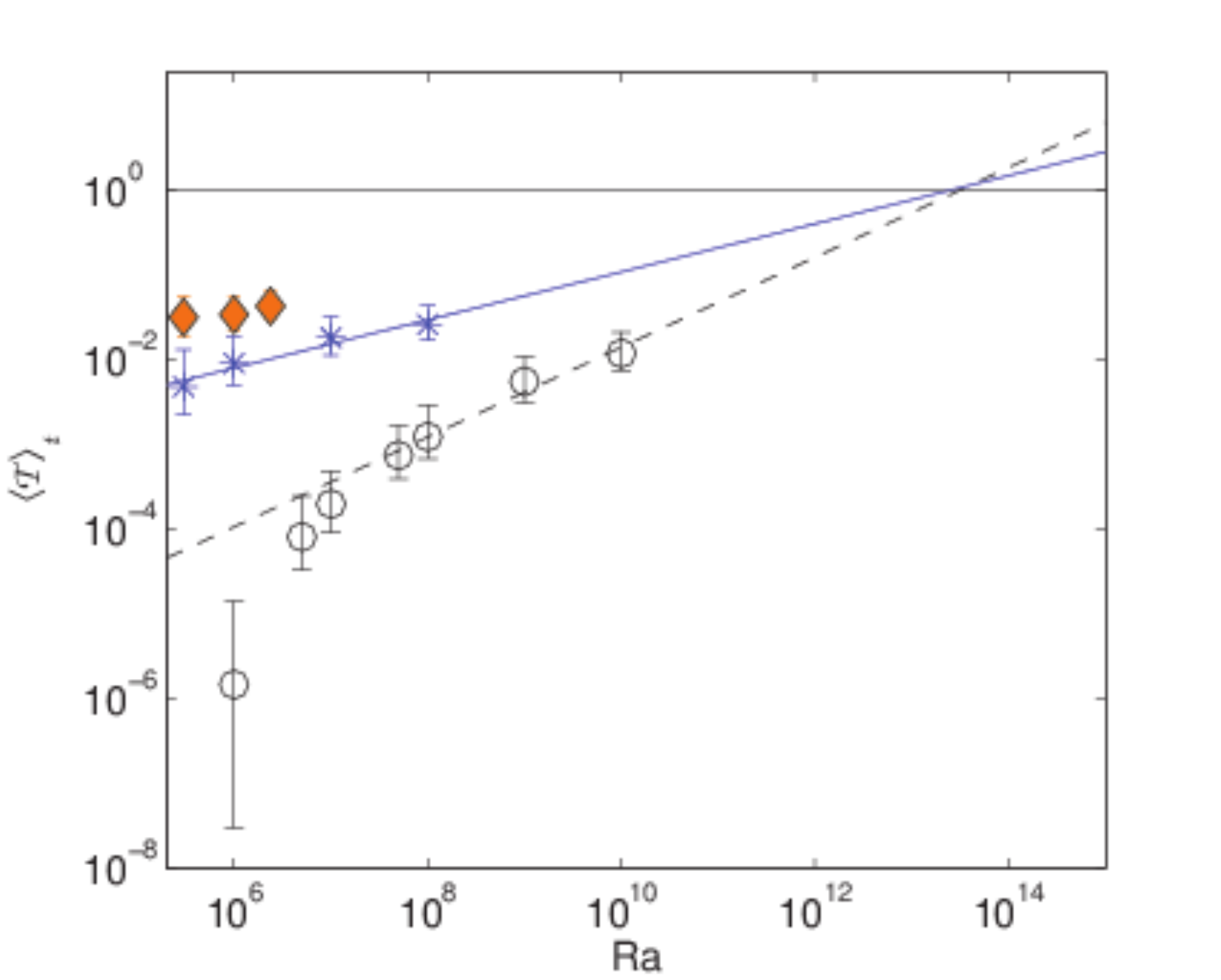}
\caption{(Colour online) Time averaged turbulent fraction $\langle {\cal T}\rangle_t$ of top and bottom plate area whose shear Reynolds number defined in (\ref{Reshloc}) is 
greater than 350 (black circles, $Pr=0.7$, magenta stars, $Pr=0.021$, yellow filled diamonds with black outline, $Pr=0.005$). The error bars correspond to the cases where $Re_{sh}^{c} = 250$ (upper) and $Re_{sh}^{c} = 450$ (lower). 
Note this fraction has been averaged over many statistically  independent time slices. The thick  black line corresponds to 
${\cal T}=1$. The fit to the last four data points for $Pr=0.7$ is  $\langle{\cal T}\rangle_t = (7 \pm 1 \times 10^{-7})Ra^{0.5\pm 0.1}$ (dashed line).
The fit to all data points for $Pr=0.021$, is $\langle {\cal T}\rangle_t = (1.6 \pm 1 \times 10^{-4})Ra^{0.28\pm 0.05}$ (solid line).}
\label{resh}
\end{figure}

Finally we plot the time averaged ${\cal T}$ in figure \ref{resh} and see that the turbulent fraction increases as a function of Rayleigh number 
for all Prandtl numbers, with the turbulent fraction always being larger for the lower Prandtl number. 

In figure \ref{resh} we also can determine the critical Rayleigh number  $Ra_{Ac}$ where $\langle {\cal T}\rangle_t=1$ for our Prandtl numbers and  $Re_{sh}^c$ 
values, which would indicate a transition to a fully turbulent  boundary layer. We find, using $Re_{sh}^c$=350 as our mid-range value, and  
$Re_{sh}^c$=250, 450 as upper and lower bounds, that $Ra_{Ac} = (3\pm 2)\times 10^{13}$ for $Pr=0.7$,  $Ra_{Ac} = (2\pm 10)\times 10^{13}$
for $Pr=0.021$. We did not fit the data or estimate an $R_{Ac}$ for $Pr=0.005$ since the data covers an interval which is too small and the Rayleigh 
numbers are too low for a reasonable extrapolation.

The uncertainty is high in these predicted critical Rayleigh numbers, since we are using data at Rayleigh numbers that are three orders of 
magnitude smaller than the predicted transition values. However, the critical Rayleigh number for $Pr=0.7$ is close to the value of 
$Ra_1^* \simeq 2\times 10^{13}$ found by \cite{He2012, Ahlers2012} for similar parameters. In these papers the authors noted tha $Ra_1^*$ 
indicated a transition from the classical laminar scaling of Nusselt number with Rayleigh number, but the fully turbulent scaling of  Nusselt 
number with Rayleigh number was not seen until $Ra_2^* \geq 5\times 10^{14}$.

Even though $\langle{\cal T}\rangle_t$ increases as $Pr$ decreases, we still find  $Ra_{Ac}$ occurs at about the same Rayleigh number for $Pr=0.021$, 
when the uncertainty is taken into account. This is because the growth of $\langle{\cal T}\rangle_t$ with Rayleigh number is so much smaller for the lower 
$Pr$. The theoretically predicted trend  by \cite{Grossmann2000} is for $Ra_{Ac}$ to decrease as  $Pr$ decreases, and the values as 
estimated from figure 2 of \cite{Grossmann2000} are 
around $10^{13},10^{14}, 10^{16}$ for $Pr=0.005, 0.021, 0.7$, respectively.
This disagreement may be due to the higher uncertainty in 
these data points, or it may be indicative of something more fundamental, i.e. that  the transition from a laminar to a turbulent boundary 
layer may be more universal than previously predicted. However, more data at higher Rayleigh numbers would provide a more precise value for $Ra_{Ac}$.

\begin{figure}
\centering
\includegraphics[width=0.8\textwidth]{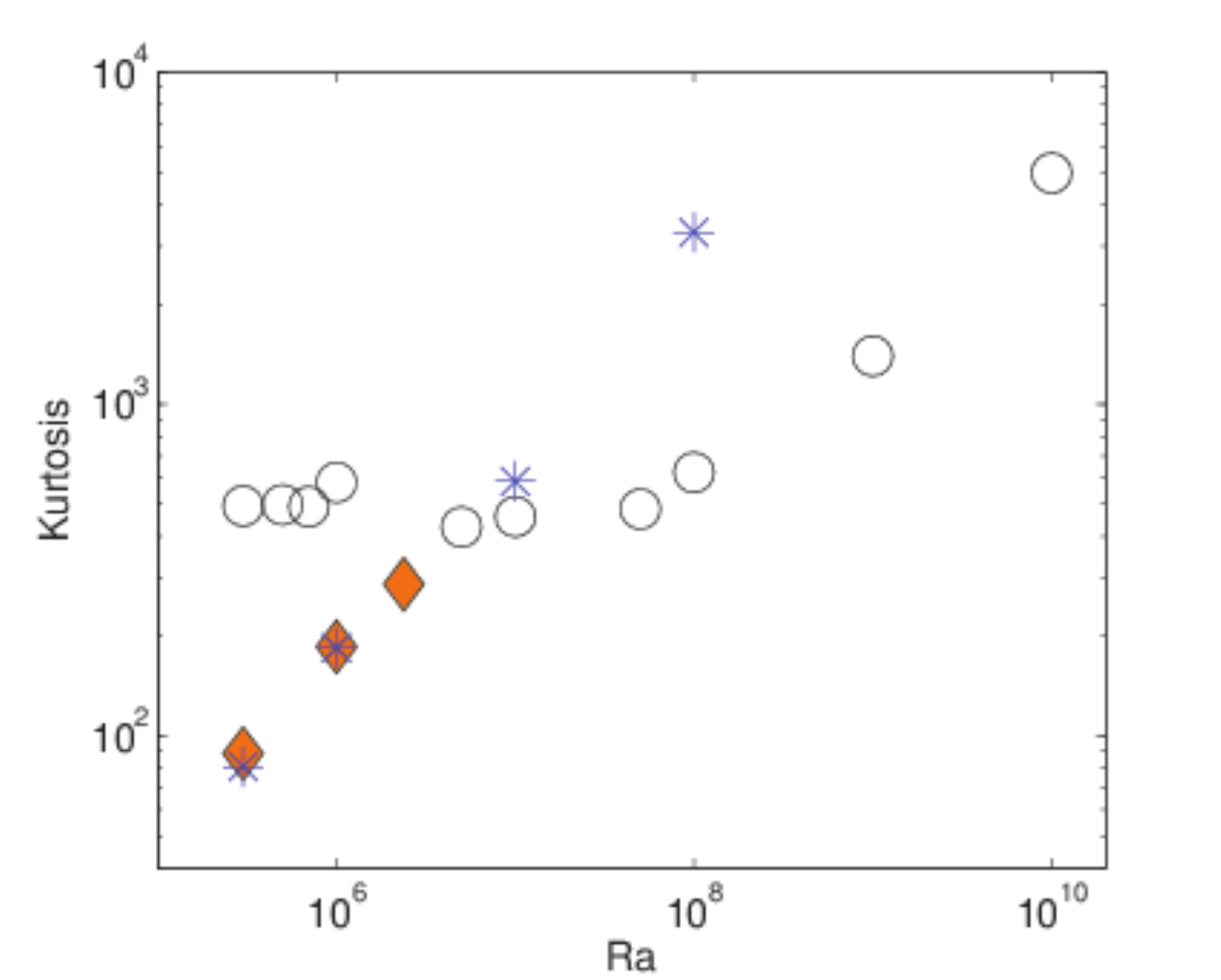}
\caption{(Colour online) Kurtosis (\ref{kurte}) versus Rayleigh number for $Pr=0.7$ (circles), $Pr=0.021$ (stars) and 
$Pr=0.005$ (diamonds).}
\label{kurt}
\end{figure}

We can also check for intermittency by evaluating the kurtosis, as defined in equation (\ref{kurte}). For a Gaussian distribution the kurtosis should be equal to three. Anything larger than this indicates wider tails to the distribution, and hence intermittency. The kurtosis is plotted in figure \ref{kurt} for all Rayleigh and Prandtl numbers. We see that the kurtosis is much larger than three for all data. For $Pr = 0.7$, the kurtosis stays fairly level at 500 until $Ra = 1\times 10^8$ and then it increases with $Ra$. However, for the lower $Pr$ data the kurtosis increases with $Ra$ for all data and the kurtosis values are also very close to one another. 

\begin{equation}
\label{kurte}
\mbox{Kurtosis} = \dfrac{\langle (Re_{sh}-\langle Re_{sh}\rangle)^4\rangle}{\langle (Re_{sh}-\langle Re_{sh}\rangle)^2\rangle^2}
\end{equation}
\section{Summary and discussion}
A comparison of the global and local statistics has been presented for low $Pr$ numbers ($Pr=0.005, 0.021$) and those of air ($Pr=0.7$) with 
very finely-resolved DNS. The low $Pr$ cases are in a parameter regime that has not been studied numerically at such fine resolution.  
It is found that the temperature field is more diffuse while the velocity field is more vigorous as the Prandtl number is lowered. It is also found that 
as the Prandtl number decreases there is a slight overall decrease in the value of the scaling exponents of diagnostic quantities, such as Nusselt 
number and thermal boundary layer thicknesses, with Rayleigh number. We attribute this to the more diffuse temperature field and resulting 
reduced heat transport as $Pr$ decreases to these very low values.

The global heat transfer scaling exponents are found to have values of $0.29 \pm 0.01$, $0.26 \pm 0.01$ and 
$0.265 \pm 0.01$ for $Pr = 0.7, 0.021$ and $0.005$ respectively. When compared  with experimental results of \cite{Horanyi1999}, 
the present exponents are found to agree quite well for $Pr=0.005$. The exponent for $Pr=0.021$ agrees with the experiments by \cite{Rossby1969},
\cite{Cioni1996}, \cite{Takeshita1996}, \cite{King2013}. However our exponent is on the high end of the range, and disagrees with the theory which predicts a scaling exponent of 1/5 for this parameter regime in \cite{Grossmann2000}.
The global momentum scaling exponents are $0.49 \pm 0.01, 0.44 \pm 0.01, 0.49 \pm 0.01$ for $Pr=0.7, 0.021, 0.005$. For the case of $Pr=0.021$, they agree well with the experimental results by \cite{Takeshita1996}. Finally it is found that when the Nusselt number is plotted
versus Prandtl number, the agreement with the Grossmann-Lohse theory is quite good, except for a slight disagreement in the lower Prandtl numbers ($0.021, 0.005$). It is stressed once more that the number of data points of our DNS is small and that the predictions
of the scaling exponents for global heat and mass transfer should be considered with caution.

The root-mean-square profiles of velocity show a significant enhancement of velocity with decreasing $Pr$ and a shift of the maxima towards 
the center of the container. The root-mean-square profiles of temperature show similar values for the location of the maxima for $Pr=0.7, 0.021, 0.005$, but 
a reduction in overall root-mean-square temperature for $Pr=0.005$. Finally we see that the magnitude of the vertical mean 
profiles of $\epsilon$ and $\epsilon_T$ systematically increases as $Pr$ decreases as is expected. The kinetic dissipation rate increases due 
to the enhanced velocity fluctuations and the thermal dissipation rate increases due to the higher diffusivity of the temperature field. The agreement 
between profiles of $\epsilon(z)$ and $\Omega(z)/\sqrt{Gr}$ are also very good, similar to that found for homogeneous 
isotropic turbulence.

When we compute the thermal boundary layer thicknesses we find that they are also consistently larger as $Pr$ decreases, as expected. Also 
the magnitude of the scaling exponents decrease slightly as $Pr$ decreases, consistent with the $Nu$ scaling. In contrast, the velocity boundary layer thicknesses 
are consistently smaller as $Pr$ decreases and the scaling exponent magnitudes increase as $Pr$ decreases. The effect of the side walls is found to remain 
significant for the lower $Pr$ cases ($0.021$ and $0.005$) as Rayleigh number increases but it becomes less significant for $Pr = 0.7$. 
  
Turning now to the local statistics, it is found that the skin-friction coefficient $c_\tau$ is significantly reduced as $Pr$ decreases for all Rayleigh 
numbers, averaging either over the central region or the whole plate. It is also found that the scaling exponent with Reynolds number when the side-wall regions are included  increases from $-0.54\pm 0.01$ 
for $Pr=0.7$ (consistent with a laminar boundary layer) to $-0.23 \pm 0.04$ for the lowest $Pr$ which bears a closer resemblance to an intermittently 
turbulent boundary layer. Inspired by this, we compute a local shear Reynolds number and resulting turbulent fraction $\langle {\cal T}\rangle_t$. 
We find that this turbulent fraction increases with Rayleigh number and is larger for the lower Prandtl numbers. By extrapolating to $\langle{\cal T}\rangle_t=1$ 
we can predict the Rayleigh number $Ra_{Ac}$ at which the boundary layer would become fully turbulent. We find  $Ra_{Ac} = (3\pm 2)\times 
10^{13}$ for $Pr=0.7$ and $Ra_{Ac} = (2\pm 10)\times 10^{13}$ for $Pr=0.021$. It is stressed once more that our prediction is based on an 
extrapolation which might be wrong once the gap to the high Rayleigh numbers is filled with further DNS data points.

The present analysis has confirmed the results of \cite{Schumacher2015}, namely that the fluid turbulence in low-$Pr$ 
convection is much more strongly driven and reaches a more vigorous turbulence level than comparable convection flows in 
air or water. The reason for this observation is that the bottleneck in the RBC system -- the thermal boundary layer -- is thicker 
due to enhanced diffusion. Coarser thermal plumes can inject kinetic energy at a larger scale as shown in \cite{Schumacher2015} 
and extend the Kolmogorov-like cascade range in the bulk. This is in line with an enhanced energy flux in the inertial range and 
a smaller Kolmogorov length. Coarser plumes intensify also the large-scale circulation in the cell and thus amplify the transient 
and intermittent behavior in the velocity boundary layer which was shown here, e.g. by the skin friction or the local shear Reynolds 
number. Our current studies indicate that low Prandtl number convection at the same Grashof number may  be more susceptible 
to a transition to turbulence in the tiny boundary layer when compared with convection in air.

Given the very low-$Pr$ data records, a next step would be to refine the analysis in the transient boundary layers and to develop alternative 
measures that quantify the increasing number of turbulent patches in the velocity boundary layer. These studies are currently under way and will be
reported elsewhere.

Helpful discussions with Jonathan Aurnou, Ronald du Puits, Peter Frick and Bruno Eckhardt are acknowledged. The work is supported by Research Unit 
1182 and the Research Training Group 1567 of the Deutsche Forschungsgemeinschaft. We acknowledge supercomputing time at the Blue 
Gene/Q JUQUEEN at the J\"ulich Supercomputing Centre which was provided by grant HIL09 of the John von Neumann Institute 
for Computing. Furthermore, we acknowledge an award of computer time provided by the INCITE program. This research used resources of the 
ALCF at ANL, which is supported by the DOE under contract DE-AC02-06CH11357.

\end{document}